\documentclass[12pt]{article}
\usepackage{epsfig}
\usepackage{amsmath}
\usepackage{graphicx}
\usepackage[colorlinks=true,citecolor=blue]{hyperref}
\usepackage{cite}

\begin{document}

\author{S. Dev\thanks{dev5703@yahoo.com} $^{a, b}$, Radha Raman Gautam\thanks{gautamrrg@gmail.com} $^a$ and Lal Singh\thanks{lalsingh96@yahoo.com} $^a$}
\title{Neutrino Mass Matrices with Two Equalities Between the Elements or Cofactors}
\date{$^a$\textit{Department of Physics, Himachal Pradesh University, Shimla 171005, INDIA.}\\
\smallskip
$^b$\textit{Department of Physics, School of Sciences, HNBG Central University, Srinagar, Uttarakhand 246174, INDIA.}}

\maketitle
\begin{abstract}
We study the implications of the existence of two equalities between the elements or cofactors of the neutrino mass matrix. There are sixty five structures of this type for each case. Phenomenological implications for unknown parameters like the effective Majorana mass of the electron neutrino and CP-violating phases are examined for the viable cases. To illustrate how such forms of the neutrino mass matrices may be realized, we also present a simple $A_4$ model for one of the classes in each case.
\end {abstract}

\section{Introduction}
After the discovery of neutrino oscillations, there has been tremendous progress in determining neutrino masses and mixings. Recently, the last unknown mixing angle $(\theta_{13})$ has been measured rather precisely\cite{t2k, minos, dchooz, dayabay, reno} and its relatively large value has, also, provided an opportunity for the measurement of CP violating phase $\delta$ in the lepton mixing matrix. The rather large value of $\theta_{13}$ has also forced modifications of models like Tribimaximal(TBM) mixing\cite{tbm} which predict $\theta_{13} = 0$, by considering deviations from TBM mixing. On the other hand, schemes like zero textures\cite{zerotexture,xingzt2011}, vanishing minors\cite{zerominor} and hybrid textures\cite{hybrid} which do not predict definite values of mixing angles but do induce relations between neutrino masses and mixing angles, readily accommodate  a non-zero and relatively large value of $\theta_{13}$. 
However, the currently available data on neutrino masses and mixings is insufficient for an unambiguous reconstruction of the neutrino mass matrix. The existing data cannot, without some additional assumptions, determine all the elements of the Yukawa coupling matrices for the neutrinos. Thus, theoretical ideas such as zero textures\cite{zerotexture,xingzt2011}, vanishing minors\cite{zerominor} and hybrid textures\cite{hybrid} which restrict the structure of the neutrino mass matrix ($M_\nu$) are important for guiding future searches.\\
In this work, we systematically study the implications of the existence of two equalities between elements (TEE) of the neutrino mass matrix or two equalities between cofactors (TEC) of the neutrino mass matrix. There are sixty five such possibilities in each case which have been listed in Table 1. We find that two equalities between the elements of $M_\nu$ can be obtained through type-II seesaw mechanism\cite{seesaw2} whereas two equalities between cofactors of $M_\nu$ arise from type-I seesaw mechanism\cite{seesaw1}.\\
In the framework of type-I seesaw mechanism, the effective Majorana neutrino mass matrix is given by 
\begin{equation}
M_\nu = -M_D M_R^{-1} M_D^T
\end{equation}
where $M_D$ is the Dirac neutrino mass matrix and $M_R$ is the right-handed Majorana mass matrix. In the diagonal basis for $M_D$, the zero textures of $M_R$ show as zero minors in $M_\nu$\cite{zerominor}. Here, we consider the possibility that $M_R$ has two equal elements and $M_D$ is proportional to the unit matrix. Such $M_D$ and $M_R$ will give rise to the neutrino mass matrix $M_\nu$ which has two equalities between the cofactors of $M_\nu$ and the two equalities of cofactors in $M_\nu$ are corresponding to the equal elements in $M_R$. Two equalities between cofactors in other words can be seen as two equalities between elements of the inverse of $M_\nu$. Thus, effectively we are studying all the possible cases of two equalities between the elements of $M_\nu$ and $M_\nu^{-1}$. The condition of two equalities between the elements or cofactors of $M_\nu$ imply constraints on the parameters of the neutrino sector which leads to a restricted parameter space for these observables. To demonstrate how such forms of $M_\nu$ can be realized, we also present a simple $A_4$ model for one of the texture structures in each case. 

\begin{table}[!]
\begin{small}
\noindent\makebox[\textwidth]{
\begin{tabular}{|c|c|c|c|c|c|c|}
\hline  & $A$ & $B$ & $C$ & $D$ & $E$ & $F$  \\
\hline $I$ & $\left(
\begin{array}{ccc}
a & a & a \\  a & d & e \\ a & e & f
\end{array}
\right)$ & $\left(
\begin{array}{ccc}
a & b & a \\  b & a & e \\ a & e & f
\end{array}
\right)$ & $\left(
\begin{array}{ccc}
a & b & c \\  b & a & a \\ c & a & f
\end{array}
\right)$ & $\left(
\begin{array}{ccc}
a & b & c \\  b & d & a \\ c & a & a
\end{array}
\right)$&$\left(
\begin{array}{ccc}
a & b & b \\  b & d & e \\ b & e & a
\end{array}
\right)$& $\left(
\begin{array}{ccc}
a & b & b \\  b & b & e \\ b & e & f
\end{array}
\right)$\\

\hline $II$ & $\left(
\begin{array}{ccc}
a & a & c \\  a & a & e \\ c & e & f
\end{array}
\right)$ & $\left(
\begin{array}{ccc}
a & b & a \\  b & d & a \\ a & a & f
\end{array}
\right)$ & $\left(
\begin{array}{ccc}
a & b & c \\  b & a & e \\ c & e & a
\end{array}
\right)$  & $\left(
\begin{array}{ccc}
a & b & b \\  b & d & a \\ b & a & f
\end{array}
\right)$ & $\left(
\begin{array}{ccc}
a & b & c \\  b & b & e \\ c & e & a
\end{array}
\right)$ & $\left(
\begin{array}{ccc}
a & b & b \\  b & d & b \\ b & b & f
\end{array}
\right)$ \\

\hline $III$ & $\left(
\begin{array}{ccc}
a & a & c \\  a & d & a \\ c & a & f
\end{array}
\right)$ & $\left(
\begin{array}{ccc}
a & b & a \\  b & d & e \\ a & e & a
\end{array}
\right)$  & $\left(
\begin{array}{ccc}
a & b & b \\  b & a & e \\ b & e & f
\end{array}
\right)$ & $\left(
\begin{array}{ccc}
a & b & c \\  b & b & a \\ c & a & f
\end{array}
\right)$  & $\left(
\begin{array}{ccc}
a & b & c \\  b & d & b \\ c & b & a
\end{array}
\right)$ & $\left(
\begin{array}{ccc}
a & b & b \\  b & d & e \\ b & e & b
\end{array}
\right)$ \\

\hline $IV$ & $\left(
\begin{array}{ccc}
a & a & c \\  a & d & e \\ c & e & a
\end{array}
\right)$ & $\left(
\begin{array}{ccc}
a & b & a \\  b & b & e \\ a & e & f
\end{array}
\right)$ &$\left(
\begin{array}{ccc}
a & b & c \\  b & a & b \\ c & b & f
\end{array}
\right)$& $\left(
\begin{array}{ccc}
a & b & c \\  b & d & a \\ c & a & b
\end{array}
\right)$ &$\left(
\begin{array}{ccc}
a & b & c \\  b & c & e \\ c & e & a
\end{array}
\right)$ & $\left(
\begin{array}{ccc}
a & b & b \\  b & d & d \\ b & d & f
\end{array}
\right)$\\

\hline $V$ & $\left(
\begin{array}{ccc}
a & a & c \\  a & c & e \\ c & e & f
\end{array}
\right)$  & $\left(
\begin{array}{ccc}
a & b & a \\  b & d & b \\ a & b & f
\end{array}
\right)$ &$\left(
\begin{array}{ccc}
a & b & c \\  b & a & e \\ c & e & b
\end{array}
\right)$  & $\left(
\begin{array}{ccc}
a & b & c \\  b & c & a \\ c & a & f
\end{array}
\right)$ & $\left(
\begin{array}{ccc}
a & b & c \\  b & d & c \\ c & c & a
\end{array}
\right)$ & $\left(
\begin{array}{ccc}
a & b & b \\  b & d & e \\ b & e & d
\end{array}
\right)$ \\

\hline $VI$ & $\left(
\begin{array}{ccc}
a & a & c \\ a & d & c \\ c & c & f
\end{array}
\right)$ & $\left(
\begin{array}{ccc}
a & b & a \\  b & d & e \\ a & e & b
\end{array}
\right)$ &$\left(
\begin{array}{ccc}
a & b & c \\  b & a & c \\ c & c & f
\end{array}
\right)$  &$\left(
\begin{array}{ccc}
a & b & c \\  b & d & a \\ c & a & c
\end{array}
\right)$ & $\left(
\begin{array}{ccc}
a & b & c \\  b & d & d \\ c & d & a
\end{array}
\right)$ &$\left(
\begin{array}{ccc}
a & b & b \\  b & d & e \\ b & e & e
\end{array}
\right)$  \\

\hline $VII$ & $\left(
\begin{array}{ccc}
a & a & c \\ a & d & e \\ c & e & c
\end{array}
\right)$ & $\left(
\begin{array}{ccc}
a & b & a \\  b & d & d \\ a & d & f
\end{array}
\right)$ & $\left(
\begin{array}{ccc}
a & b & c \\  b & a & e \\ c & e & c
\end{array}
\right)$&$\left(
\begin{array}{ccc}
a & b & c \\  b & d & a \\ c & a & d
\end{array}
\right)$ & $\left(
\begin{array}{ccc}
a & b & c \\  b & d & b \\ c & b & b
\end{array}
\right)$ & $\left(
\begin{array}{ccc}
a & b & c \\  b & b & b \\ b & b & f
\end{array}
\right)$  \\

\hline $VIII$ & $\left(
\begin{array}{ccc}
a & a & c \\ a & d & d \\ c & d & f
\end{array}
\right)$ & $\left(
\begin{array}{ccc}
a & b & a \\  b & d & e \\ a & e & d
\end{array}
\right)$ &$\left(
\begin{array}{ccc}
a & b & c \\  b & a & e \\ a & e & e
\end{array}
\right)$  & $\left(
\begin{array}{ccc}
a & b & c \\  b & c & e \\ c & e & b
\end{array}
\right)$ & $\left(
\begin{array}{ccc}
a & b & c \\  b & c & b \\ c & b & f
\end{array}
\right)$ & $\left(
\begin{array}{ccc}
a & b & c \\  b & b & e \\ c & e & b
\end{array}
\right)$  \\

\hline $IX$ & $\left(
\begin{array}{ccc}
a & a & c \\ a & d & e \\ c & e & d
\end{array}
\right)$ & $\left(
\begin{array}{ccc}
a & b & a \\  b & d & e \\ a & e & e
\end{array}
\right)$&$\left(
\begin{array}{ccc}
a & b & c \\  b & c & c \\ c & c & f
\end{array}
\right)$ & $\left(
\begin{array}{ccc}
a & b & c \\  b & d & c \\ c & c & b
\end{array}
\right)$ & $\left(
\begin{array}{ccc}
a & b & c \\  b & d & b \\ c & b & c
\end{array}
\right)$ & $\left(
\begin{array}{ccc}
a & b & c \\  b & b & c \\ c & c & f
\end{array}
\right)$  \\

\hline $X$ & $\left(
\begin{array}{ccc}
a & a & c \\ a & d & e \\ c & e & e
\end{array}
\right)$ &$\left(
\begin{array}{ccc}
a & b & c \\  b & d & c \\ c & c & c
\end{array}
\right)$&$\left(
\begin{array}{ccc}
a & b & c \\  c & c & e \\ c & e & c
\end{array}
\right)$ &$\left(
\begin{array}{ccc}
a & b & c \\  b & d & d \\ c & d & b
\end{array}
\right)$ & $\left(
\begin{array}{ccc}
a & b & c \\  b & d & b \\ c & b & d
\end{array}
\right)$ & $\left(
\begin{array}{ccc}
a & b & c \\  b & b & e \\ c & e & c
\end{array}
\right)$  \\

\hline $XI$ & - & $\left(
\begin{array}{ccc}
a & b & c \\  b & d & c \\ c & c & d
\end{array}
\right)$ & $\left(
\begin{array}{ccc}
a & b & c \\  b & c & e \\ c & e & e
\end{array}
\right)$ & $\left(
\begin{array}{ccc}
a & b & c \\  b & d & d \\ c & d & d
\end{array}
\right)$ & $\left(
\begin{array}{ccc}
a & b & c \\  b & d & d \\ c & d & c
\end{array}
\right)$ & $\left(
\begin{array}{ccc}
a & b & c \\  b & b & e \\ c & e & e
\end{array}
\right)$  \\
\hline
\end{tabular}}
\end{small}
\caption{Sixty Five possible texture structures of $M_{\nu}$ or $M_{R}$ with two equalities.}
\end{table} 
\section{Formalism}
We reconstruct the neutrino mass matrix in the flavor basis assuming neutrinos to be Majorana particles. In this basis, a complex symmetric neutrino mass matrix can be diagonalized by a unitary matrix $V'$ as
\begin{equation}
M_{\nu}= V' M_{\nu}^{diag}V'^{T}
\end{equation}
where $M_{\nu}^{diag}$ = diag$(m_1,m_2,m_3)$. \\
The unitary matrix $V'$ can be parametrized as
\begin{equation}
V' = P_lV \ \ \ \textrm{with}\ \ \ \ V = UP_\nu
\end{equation}
where  \cite{foglipdg}
\begin{equation}
U= \left(
\begin{array}{ccc}
c_{12}c_{13} & s_{12}c_{13} & s_{13}e^{-i\delta} \\
-s_{12}c_{23}-c_{12}s_{23}s_{13}e^{i\delta} &
c_{12}c_{23}-s_{12}s_{23}s_{13}e^{i\delta} & s_{23}c_{13} \\
s_{12}s_{23}-c_{12}c_{23}s_{13}e^{i\delta} &
-c_{12}s_{23}-s_{12}c_{23}s_{13}e^{i\delta} & c_{23}c_{13}
\end{array}
\right)
\end{equation} with $s_{ij}=\sin\theta_{ij}$ and $c_{ij}=\cos\theta_{ij}$ and
\begin{small}
\begin{center}
$P_\nu = \left(
\begin{array}{ccc}
1 & 0 & 0 \\ 0 & e^{i\alpha} & 0 \\ 0 & 0 & e^{i(\beta+\delta)}
\end{array}
\right)$ , \ \ \ 
$P_l = \left(
\begin{array}{ccc}
e^{i\varphi_e} & 0 & 0 \\ 0 & e^{i\varphi_\mu} & 0 \\ 0 & 0 & e^{i\varphi_\tau}
\end{array}
\right).$
\end{center}
\end{small}
$P_\nu$ is the diagonal phase matrix with
the two Majorana-type CP- violating phases $\alpha$, $\beta$ and one Dirac-type CP-violating phase $\delta$. The phase matrix $P_l$ is unphysical and depends on the phase convention. The matrix $V$ is called the neutrino mixing matrix or the Pontecorvo-Maki-Nakagawa-Sakata (PMNS) matrix  \cite{pmns}. Using Eq. (2) and Eq. (3), the neutrino mass matrix can be written as
\begin{equation}
M_{\nu}=P_l U P_\nu M_{\nu}^{diag}P_\nu^{T}U^{T}P_l^T.
\end{equation}
The CP violation in neutrino oscillation experiments can be described through a rephasing invariant quantity, $J_{CP}$ \cite{jarlskog} with $J_{CP}=Im(U_{e1}U_{\mu2}U_{e2}^*U_{\mu1}^*)$. In the above parametrization, $J_{CP}$ is given by
\begin{equation}
J_{CP} = s_{12}s_{23}s_{13}c_{12}c_{23}c_{13}^2 \sin \delta \   .
\end{equation}

\subsection{Two Equalities Between the Elements of $M_\nu$}
The simultaneous existence of two equalities between the elements of the neutrino mass matrix implies
\begin{align}
& e^{i(\varphi_a +\varphi_b)}M_{\nu (ab)} - e^{i(\varphi_c +\varphi_d)}M_{\nu (cd)} = 0\ , \\ & e^{i(\varphi_{a'}+\varphi_{b'})} M_{\nu (a'b')} - e^{i(\varphi_{c'} +\varphi_{d'})} M_{\nu (c'd')} = 0 
\end{align}
or
\begin{align}
& Q M_{\nu (ab)} - M_{\nu (cd)} = 0\ , \\ & Q' M_{\nu (a'b')} - M_{\nu (c'd')} = 0 
\end{align}
where
\begin{align}
Q = & e^{i(\varphi_a +\varphi_b - (\varphi_c +\varphi_d))}\ , \\ Q' = & e^{i(\varphi_{a'}+\varphi_{b'}-(\varphi_{c'} +\varphi_{d'}))}\ .
\end{align}

These two conditions yield two complex equations viz.
\begin{align}
\sum_{i=1}^{3} & (Q V_{ai}V_{bi} - V_{ci}V_{di})m_i = 0 \ , \\
\sum_{i=1}^{3} & (Q' V_{a'i}V_{b'i} - V_{c'i}V_{d'i})m_i = 0  \ .
\end{align}
The above equations can be rewritten as
\begin{eqnarray}
m_1 A_1 + m_2 A_2 e^{2i\alpha} + m_3 A_3 e^{2i(\beta +\delta)}=0 \ , \\
m_1 B_1 + m_2 B_2 e^{2i\alpha} + m_3 B_3 e^{2i(\beta +\delta)}=0 
\end{eqnarray}
where
\begin{equation}
A_i =(Q U_{ai}U_{bi} - U_{ci}U_{di}) \ , \ \ 
B_i =(Q' U_{a'i}U_{b'i} - U_{c'i}U_{d'i}) \ 
\end{equation}
with $(i = 1, 2, 3)$. These two complex Eqs.(15) and (16) involve nine physical parameters which include $m_{1}$, $m_{2}$, $m_{3}$, $\theta _{12}$, $\theta _{23}$, $\theta _{13}$ and three CP-violating phases $\alpha $, $\beta $ and $\delta $. In addition, there are three unphysical phases $(\varphi_e, \varphi_\mu, \varphi_\tau)$ which enter in the mass ratios as two phase differences and, in some cases, as a single phase difference. The masses $m_{2}$ and $m_{3}$ can be calculated from the mass-squared
differences $\Delta m_{21}^{2}$ and $|\Delta m_{32}^{2}|$ using the relations
\begin{equation}
m_{2}=\sqrt{m_{1}^{2}+\Delta m_{21}^{2}} \ , \ \  m_{3}=\sqrt{m_{2}^{2}+|\Delta m_{32}^{2}|} \ 
\end{equation}
where $m_2 > m_3$ for an Inverted Spectrum (IS) and $m_2 < m_3$ for the Normal Spectrum (NS). 
Using the experimental inputs of the two mass-squared differences and the three mixing angles we can constrain the other parameters. Simultaneously solving Eqs.(15) and (16) for two mass ratios, we obtain
\begin{small}
\begin{equation}
\frac{m_1}{m_2}e^{-2i\alpha }=\frac{A_2 B_3 - A_3 B_2}{A_3 B_1 - A_1 B_3}
\end{equation}
\end{small}
and
\begin{small}
\begin{equation}
\frac{m_1}{m_3}e^{-2i\beta }=\frac{A_3 B_2 - A_2 B_3 }{A_2 B_1- A_1 B_2}e^{2i\delta } \ .
\end{equation}
\end{small}
The magnitudes of the two mass ratios in Eqs.(19, 20), are given by
\begin{equation}
\rho=\left|\frac{m_1}{m_3}e^{-2i\beta }\right| ,
\end{equation}
\begin{equation}
\sigma=\left|\frac{m_1}{m_2}e^{-2i\alpha }\right|
\end{equation}
 while the CP- violating Majorana phases $\alpha$ and $\beta$ are given by
 \begin{small}
\begin{align}
\alpha & =-\frac{1}{2}arg\left(\frac{A_2 B_3 - A_3 B_2}{A_3 B_1 - A_1 B_3}\right), \\
\beta & =-\frac{1}{2}arg\left(\frac{A_3 B_2 - A_2 B_3 }{A_2 B_1- A_1 B_2}e^{2i\delta }\right).
\end{align}
\end{small}
Since, $\Delta m_{21}^{2}$ and $|\Delta m_{32}^{2}|$ are known experimentally, the values of mass ratios $(\rho,\sigma)$ from Eqs.(21) and (22) can be used to calculate $m_1$.
This can be done by inverting Eqs.(21) and (22) to obtain the two values of $m_1$, viz.,
\begin{small}
\begin{equation}
m_{1}=\sigma \sqrt{\frac{ \Delta
m_{21}^{2}}{1-\sigma ^{2}}} \ , \ \ 
m_{1}=\rho \sqrt{\frac{\Delta m_{21}^{2}+
|\Delta m_{32}^{2}|}{ 1-\rho^{2}}} .
\end{equation}
\end{small}
Similar to the case of zero textures\cite{xingzt2011}, there exists a permutation symmetry between different patterns of two equalities in $M_\nu$ corresponding to the permutation in the 2-3 rows and 2-3 columns of $M_\nu$. The corresponding permutation matrix is given by
\begin{small} 
\begin{equation}
P_{23} = \left(
\begin{array}{ccc}
1&0&0\\
0&0&1\\
0&1&0\\
\end{array}
\right).
\end{equation}
\end{small}
For example, the neutrino mass matrix for class $IF$ can be obtained from class $IIIF$ by the transformation
\begin{equation}
M_{\nu}^{IF} = P_{23}M_{\nu}^{IIIF}P_{23}^T \ .
\end{equation}
This leads to the following relations between the parameters for the classes related by the permutation symmetry: 
\begin{equation}
\theta_{12}^{IF} = \theta_{12}^{IIIF}, \ \theta_{13}^{IF} = \theta_{13}^{IF}, \ \theta_{23}^{IF} = \frac{\pi}{2}-\theta_{23}^{IIIF}, \ \delta^{IF} = \delta^{IIIF} - \pi \ .
\end{equation}
The textures related by the 2-3 permutation symmetry are
\begin{small}
\begin{align}
& IB \leftrightarrow IVA, \ IC \leftrightarrow ID, \ IE \leftrightarrow IIIC, \ IF \leftrightarrow IIIF, \ IIA \leftrightarrow IIIB, \ IIB \leftrightarrow IIIA, \nonumber \\ 
& IIE \leftrightarrow VIIC, \ IIID \leftrightarrow VID, \ 
IIIE \leftrightarrow VIC, \ IVB \leftrightarrow VIIA, \ IVC \leftrightarrow VE, \ IVD \leftrightarrow VD, \nonumber  \\ 
& IVE \leftrightarrow VC, \ IVF \leftrightarrow VIF, \ VA \leftrightarrow VIB, \ VB \leftrightarrow VIA, \ VIE \leftrightarrow VIIIC, \ VIIB \leftrightarrow XA, \nonumber  \\ 
& VIIE \leftrightarrow IXC, \ VIIF \leftrightarrow XB, \ VIIIA \leftrightarrow IXB, \ VIIIB \leftrightarrow IXA, \ VIIIE \leftrightarrow IXD,  \nonumber \\ 
& VIIIF \leftrightarrow XC, \ IXE \leftrightarrow IXF, \ XD \leftrightarrow XIC, \ XE \leftrightarrow XIB, \ XIE \leftrightarrow XIF. 
\end{align}
\end{small}
The remaining textures
\begin{small}
\begin{align}
IA, \ IIC, \ IID, \ IIF, \ VF, \ VIID, \ VIIID, \ XF, \ XID
\end{align}
\end{small}
transform unto themselves. It is interesting to note that class $VF$ is the widely studied $\mu - \tau$ symmetric texture structure \cite{mutau}.

\subsection{Two Equalities Between the Cofactors of $M_\nu$}
The simultaneous existence of two equalities between the cofactors of the neutrino mass matrix implies
\begin{align}
(-1^{(\gamma \xi)})( & e^{i(\varphi_a +\varphi_b + \varphi_c +\varphi_d)}M_{\nu (ab)} M_{\nu (cd)}- e^{i(\varphi_f +\varphi_g + \varphi_m +\varphi_n)}M_{\nu
(fg)} M_{\nu (mn)})- \\ \nonumber (-1^{(\zeta \eta)})( & e^{i(\varphi_p +\varphi_q + \varphi_r +\varphi_s)}M_{\nu (pq)} M_{\nu (rs)}- e^{i(\varphi_t +\varphi_u + \varphi_v +\varphi_w)}M_{\nu
(tu)} M_{\nu (vw)})=0 \ , \\ (-1^{(\gamma' \xi')})( & e^{i(\varphi_{a'} +\varphi_{b'} + \varphi_{c'} +\varphi_{d'})}M_{\nu (a'b')} M_{\nu (c'd')}- e^{i(\varphi_{f'} +\varphi_{g'} + \varphi_{m'} +\varphi_{n'})}M_{\nu
(f'g')} M_{\nu (m'n')})- \\ \nonumber (-1^{(\zeta' \eta')})( & e^{i(\varphi_{p'} +\varphi_{q'} + \varphi_{r'} +\varphi_{s'})}M_{\nu (p'q')} M_{\nu (r's')}- e^{i(\varphi_{t'} +\varphi_{u'} + \varphi_{v'} +\varphi_{w'})}M_{\nu
(t'u')} M_{\nu (v'w')})=0 \ .
\end{align}
or
\begin{align}
&(-1^{(\gamma \xi)})( Q_1 M_{\nu (ab)} M_{\nu (cd)}- Q_2 M_{\nu
(fg)} M_{\nu (mn)})- \nonumber \\ &(-1^{(\zeta \eta)})( Q_3 M_{\nu (pq)} M_{\nu (rs)}- Q_4 M_{\nu
(tu)} M_{\nu (vw)})=0 \ , \\ &(-1^{(\gamma' \xi')})( Q'_1 M_{\nu (a'b')} M_{\nu (c'd')}- Q'_2 M_{\nu
(f'g')} M_{\nu (m'n')})- \nonumber \\ & (-1^{(\zeta' \eta')})( Q'_3 M_{\nu (p'q')} M_{\nu (r's')}- Q'_4 M_{\nu
(t'u')} M_{\nu (v'w')})=0 \ ,
\end{align}

it is inherent in the properties of cofactors that when we substitute $\varphi_j, (j = e, \mu, \tau$), $Q_1 = Q_2$, $Q_3 = Q_4$, $Q'_1 = Q'_2$ and  $Q'_3 = Q'_4$. Thus, we have
\begin{align}
&(-1^{(\gamma \xi)})Q_1( M_{\nu (ab)} M_{\nu (cd)}- M_{\nu
(fg)} M_{\nu (mn)})- \nonumber \\ & (-1^{(\zeta \eta)})Q_3( M_{\nu (pq)} M_{\nu (rs)}- M_{\nu
(tu)} M_{\nu (vw)})=0 \ , \\ &(-1^{(\gamma' \xi')})Q'_1( M_{\nu (a'b')} M_{\nu (c'd')}- M_{\nu
(f'g')} M_{\nu (m'n')})- \nonumber \\ & (-1^{(\zeta' \eta')})Q'_3( M_{\nu (p'q')} M_{\nu (r's')}- M_{\nu
(t'u')} M_{\nu (v'w')})=0 \ ,
\end{align}
or
\begin{align}
&(-1^{(\gamma \xi)})Q( M_{\nu (ab)} M_{\nu (cd)}- M_{\nu
(fg)} M_{\nu (mn)})- \nonumber \\ & (-1^{(\zeta \eta)})( M_{\nu (pq)} M_{\nu (rs)}- M_{\nu
(tu)} M_{\nu (vw)})=0 \ , \\ &(-1^{(\gamma' \xi')})Q'( M_{\nu (a'b')} M_{\nu (c'd')}- M_{\nu
(f'g')} M_{\nu (m'n')})- \nonumber \\ & (-1^{(\zeta' \eta')})( M_{\nu (p'q')} M_{\nu (r's')}- M_{\nu
(t'u')} M_{\nu (v'w')})=0 \ ,
\end{align}
where $Q = \frac{Q_1}{Q_3}$ and $Q' = \frac{Q'_1}{Q'_3}$.\\
The above two conditions take the following form when expressed in terms of mixing matrix elements and mass eigenvalues:
\begin{align}
\sum_{l,k=1}^{3} & \{(-1^{(\gamma \xi)})Q(V_{al}V_{bl}V_{ck}V_{dk}-V_{fl}V_{gl}V_{mk}V_{nk})- \nonumber \\ &(-1^{(\zeta \eta)})(V_{pl}V_{ql}V_{rk}V_{sk}-V_{tl}V_{ul}V_{vk}V_{wk})\}m_{l}m_{k}=0 \ , \\
\sum_{l,k=1}^{3} & \{(-1^{(\gamma' \xi')})Q'(V_{a'l}V_{b'l}V_{c'k}V_{d'k}-V_{f'l}V_{g'l}V_{m'k}V_{n'k})- \nonumber \\ &(-1^{(\zeta' \eta')})(V_{p'l}V_{q'l}V_{r'k}V_{s'k}-V_{t'l}V_{u'l}V_{v'k}V_{w'k})\}m_{l}m_{k}=0  \ .
\end{align}
The above equations can be rewritten as
\begin{eqnarray}
m_1 m_2 C_3e^{2i\alpha} + m_2 m_3 C_1e^{2i(\alpha+\beta +\delta )}+m_3 m_1 C_2e^{2i(\beta +\delta)}=0 \ , \\
m_1 m_2 D_3e^{2i\alpha} + m_2 m_3 D_1e^{2i(\alpha+\beta +\delta )}+m_3 m_1 D_2e^{2i(\beta +\delta)}=0 
\end{eqnarray}
where
\begin{align}
C_h & =(-1^{(\gamma \xi)})Q(U_{al}U_{bl}U_{ck}U_{dk}-U_{fl}U_{gl}U_{mk}U_{nk})- \nonumber \\ & (-1^{(\zeta \eta)})(U_{pl}U_{ql}U_{rk}U_{sk}-U_{tl}U_{ul}U_{vk}U_{wk})+(l\leftrightarrow k) \ ,\\ \nonumber
D_h & =(-1^{(\gamma' \xi')})Q'(U_{a'l}U_{b'l}U_{c'k}U_{d'k}-U_{f'l}U_{g'l}U_{m'k}U_{n'k})- \nonumber \\ & (-1^{(\zeta' \eta')})( U_{p'l}U_{q'l}U_{r'k}U_{s'k}-U_{t'l}U_{u'l}U_{v'k}U_{w'k})+(l\leftrightarrow k) 
\end{align}
with $(h,l,k)$ as the cyclic permutation of (1,2,3). Simultaneously solving Eqs.(41) and (42) for the two mass ratios, we obtain
\begin{small}
\begin{equation}
\frac{m_1}{m_2}e^{-2i\alpha }=\frac{C_3 D_1 - C_1 D_3}{C_2 D_3 - C_3 D_2}
\end{equation}
\end{small}
and
\begin{small}
\begin{equation}
\frac{m_1}{m_3}e^{-2i\beta }=\frac{C_2 D_1 - C_1 D_2 }{C_3 D_2 - C_2 D_3}e^{2i\delta } \ .
\end{equation}
\end{small}

The magnitudes of the above two mass ratios are given by
\begin{equation}
\rho=\left|\frac{m_1}{m_3}e^{-2i\beta }\right| ,
\end{equation}
\begin{equation}
\sigma=\left|\frac{m_1}{m_2}e^{-2i\alpha }\right|
\end{equation}
 while the CP- violating Majorana phases $\alpha$ and $\beta$ are given by
 \begin{small}
\begin{align}
\alpha & =-\frac{1}{2}arg\left(\frac{C_3 D_1 - C_1 D_3}{C_2 D_3 - C_3 D_2}\right), \\
\beta & =-\frac{1}{2}arg\left(\frac{C_2 D_1 - C_1 D_2 }{C_3 D_2 - C_2 D_3}e^{2i\delta }\right).
\end{align}
\end{small}
Again there exists a permutation symmetry between the different classes of two equalities of cofactors in $M_\nu$ which corresponds to the permutation in the 2-3 rows and 2-3 columns of $M_\nu$. The classification of these texture structures is similar to the case of two equalities between elements of $M_\nu$.
\section{Numerical Results}
The current experimental constraints on neutrino parameters at 1, 2 and 3$\sigma$ \cite{valledata} are given in Table 2.
The two values of $m_1$ obtained from Eq. (21) and Eq. (22) for TEE [or Eqs. (48) and Eq. (49) for TEC] must be equal to within the errors of the oscillation parameters, for the simultaneous existence of two equalities between elements or cofactors of $M_\nu$, respectively. The known oscillation parameters are varied randomly within their 3$\sigma$ experimental ranges. The unconstrained Dirac-type CP-violating phase $\delta$ is varied randomly within its full possible range. For the numerical analysis we generate $10^7$ points ( $10^8$ when the number of allowed points is small). For most of the viable cases we obtain a lower bound on the effective Majorana mass of the electron neutrino ($M_{ee}$). The observation of neutrinoless double beta (NDB) decay would signal lepton number violation and imply Majorana nature of neutrinos, for recent reviews on NDB decay see \cite{ndbdecay}.
\begin{table}[!]
\begin{center}
\begin{tabular}{|c|c|}
\hline Parameter & Mean $^{(+1 \sigma, +2 \sigma, +3 \sigma)}_{(-1 \sigma, -2 \sigma, -3 \sigma)}$ \\
\hline $\Delta m_{21}^{2} [10^{-5}eV^{2}]$ & $7.62_{(-0.19,-0.35,-0.5)}^{(+0.19,+0.39,+0.58)}$ \\ 
\hline $\Delta m_{31}^{2} [10^{-3}eV^{2}]$ & $2.55_{(-0.09,-0.19,-0.24)}^{(+0.06,+0.13,+0.19)}$, \\&
$(-2.43_{(-0.07,-0.15,-0.21)}^{(+0.09,+0.19,+0.24)})$ \\ 
\hline $\sin^2 \theta_{12}$ & $0.32_{(-0.017,-0.03,-0.05)}^{(+0.016,+0.03,+0.05)}$ \\ 
\hline $\sin^2 \theta_{23}$ & $0.613_{(-0.04,-0.233,-0.25)}^{(+0.022,+0.047,+0.067)}$, \\& $(0.60_{(-0.031,-0.210,-0.230)}^{(+0.026,+0.05,+0.07)})$ \\ 
\hline $\sin^2 \theta_{13}$ & $0.0246_{(-0.0029,-0.0054,-0.0084)}^{(+0.0028,+0.0056,+0.0076)}$,\\& $(0.0250_{(-0.0027,-0.005,-0.008)}^{(+0.0026,+0.005,+0.008)})$ \\ 
\hline 
\end{tabular}
\caption{Current Neutrino oscillation parameters from global fits \cite{valledata}. The upper (lower) row corresponds to Normal (Inverted) Spectrum, with $\Delta m^2_{31} > 0$ ($\Delta m^2_{31} < 0$).}
\end{center}
\end{table}
$M_{ee}$ which determines the rate of NDB decay is given by
\begin{equation}
M_{ee}= |m_1c_{12}^2c_{13}^2+ m_2s_{12}^2c_{13}^2 e^{2i\alpha}+ m_3s_{13}^2e^{2i\beta}|.
\end{equation}
NDB decay provides a window to probe the neutrino mass scale. Part of the Heidelberg-Moscow collaboration claimed a positive signal for NDB decay corresponding to $M_{ee} = (0.11 - 0.56)$ eV at 95$\%$ C. L. \cite{klapdormee}. However, this claim was subsequently criticized in \cite{ferugliomee}. The results reported in \cite{klapdormee} will be checked in the currently running and forthcoming NDB decay experiments. 
There are a large number of projects such as CUORICINO\cite{cuoricino}, CUORE \cite{cuore}, GERDA \cite{gerda}, MAJORANA \cite{majorana}, SuperNEMO \cite{supernemo}, EXO \cite{exo}, GENIUS \cite{genius} which aim to achieve a sensitivity upto 0.01 eV for $M_{ee}$.  In our numerical analysis, we take the upper limit of $M_{ee}$ to be 0.5 eV. Cosmological observations put an upper bound on the sum of neutrino masses
\begin{equation}
\Sigma = \sum_{i=1}^3 m_i.
\end{equation}
The nine-year WMAP data alone restrict $\Sigma$ to be less than $1.3$ eV (95\% C.L.) \cite{wmap}. The combined WMAP+eCMB+BAO+H$_0$ data limit $\Sigma$ $<$ $0.44$ eV at $95 \%$ C.L. \cite{wmap}. However, these limits strongly depend on the details of the model considered and the data set used. In our numerical analysis we have taken the upper limit of  $\Sigma$ to be 0.9 eV.
\subsection{Numerical Results for TEE}
In this section, we present the numerical results for textures with two equalities between the elements of the neutrino mass matrix. The main outcomes of our analysis are: 
\begin{figure}[!]
\begin{center}
\epsfig{file=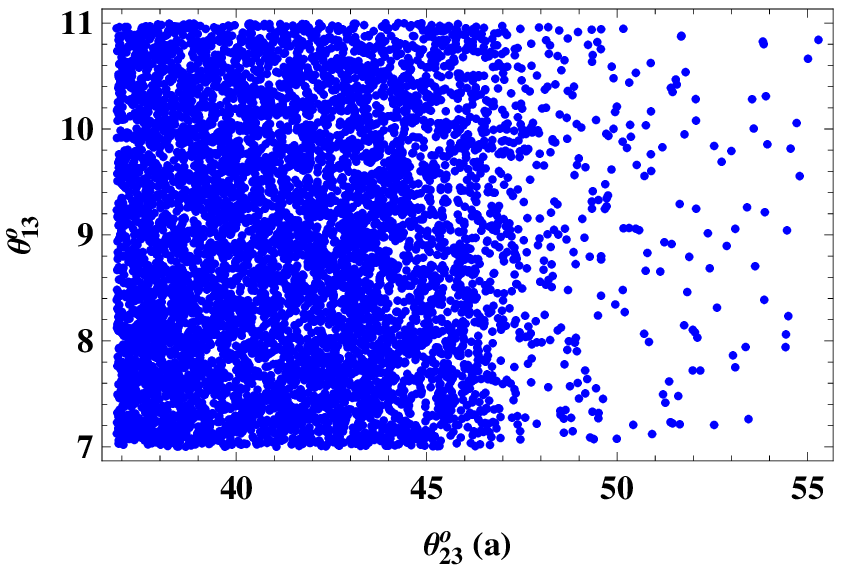,height=4.5cm,width=4.5cm}
\epsfig{file=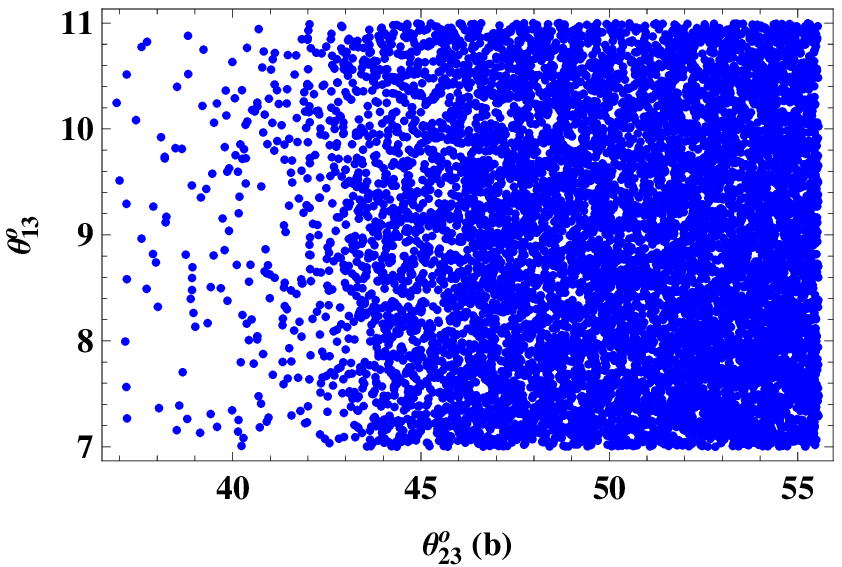,height=4.5cm,width=4.5cm}\\
\epsfig{file=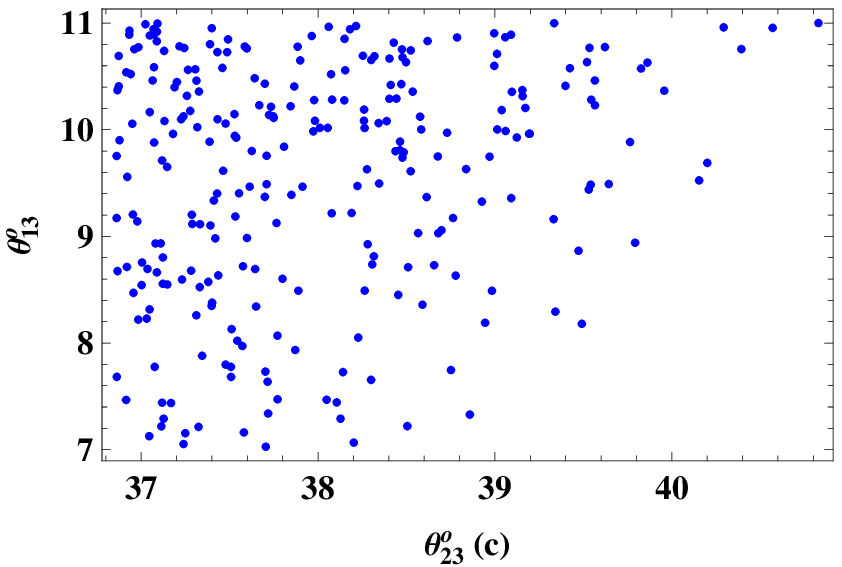,height=4.5cm,width=4.5cm}
\epsfig{file=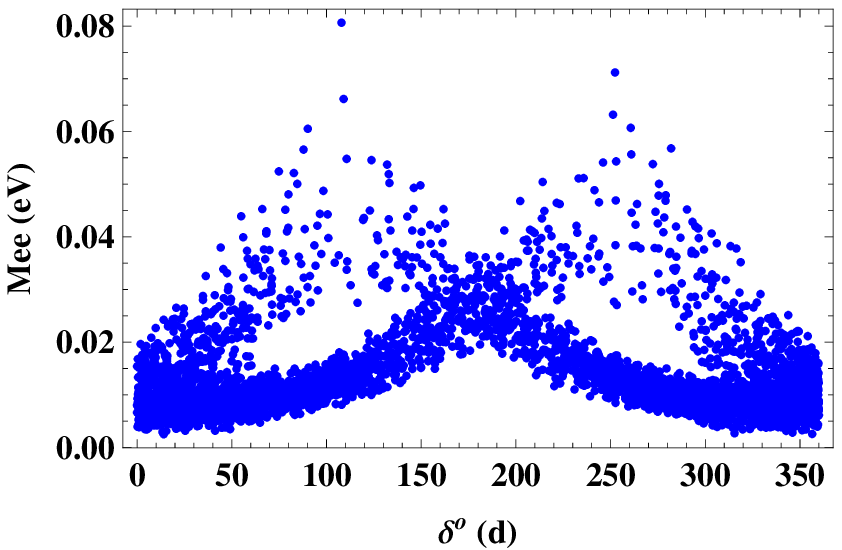,height=4.5cm,width=4.5cm}
\end{center}
\caption{The TEE correlation plots for classes $IF(a)$(NS), $IIIF(b)$(NS), $IIIB(c)$(IS) and $VB(d)$(NS).}
\end{figure}

\begin{figure}[!]
\begin{center}
\epsfig{file=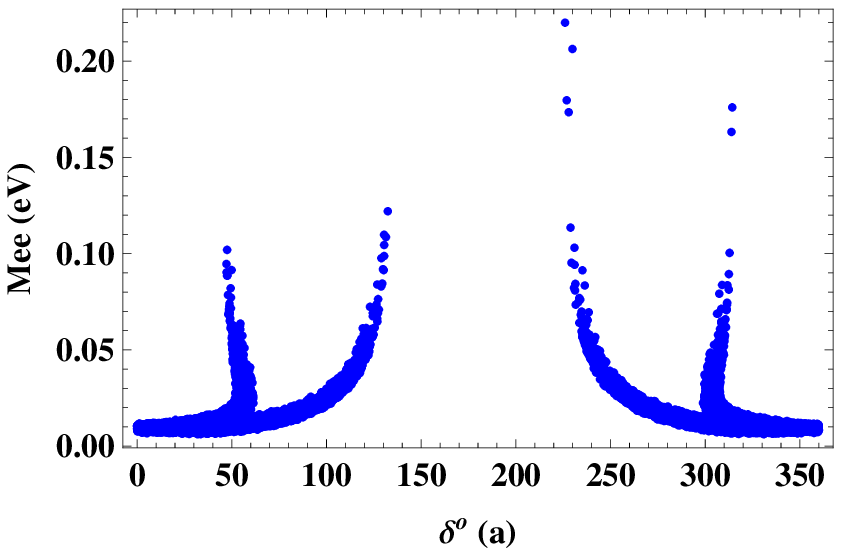,height=4.5cm,width=4.5cm}
\epsfig{file=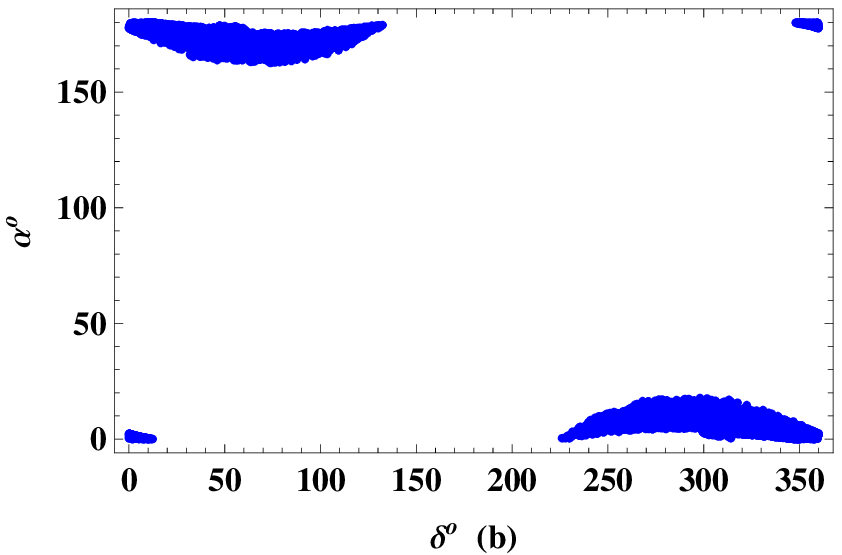,height=4.5cm,width=4.5cm}\\
\epsfig{file=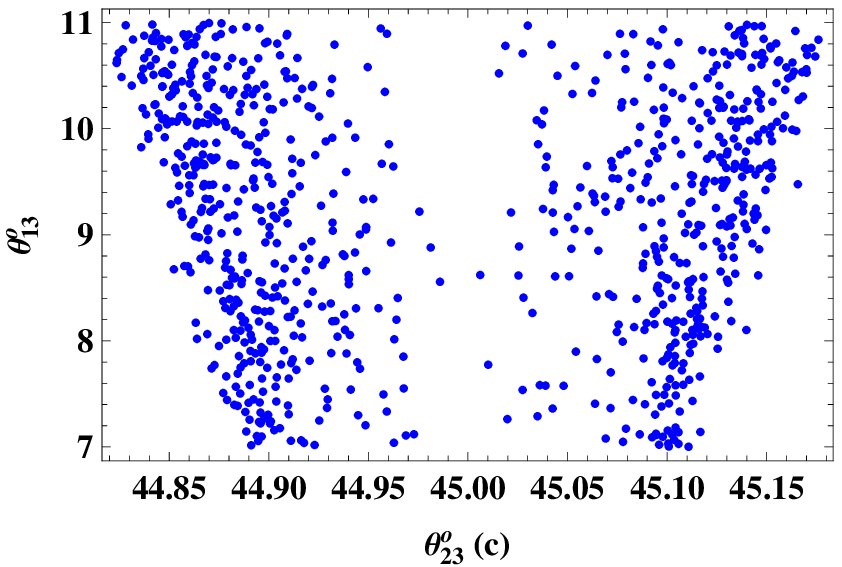,height=4.5cm,width=4.5cm}
\epsfig{file=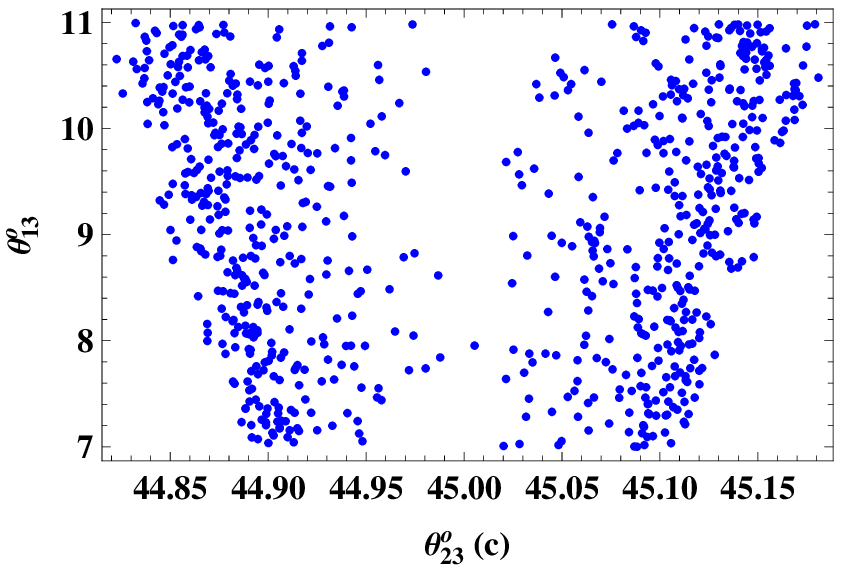,height=4.5cm,width=4.5cm}
\end{center}
\caption{The TEE correlation plots for classes $IVF(a)$(NS), $IVF(b)$(NS), $VIIID(c)$(NS) and $VIIID(d)$(IS).}
\end{figure}

\begin{figure}[!]
\begin{center}
\epsfig{file=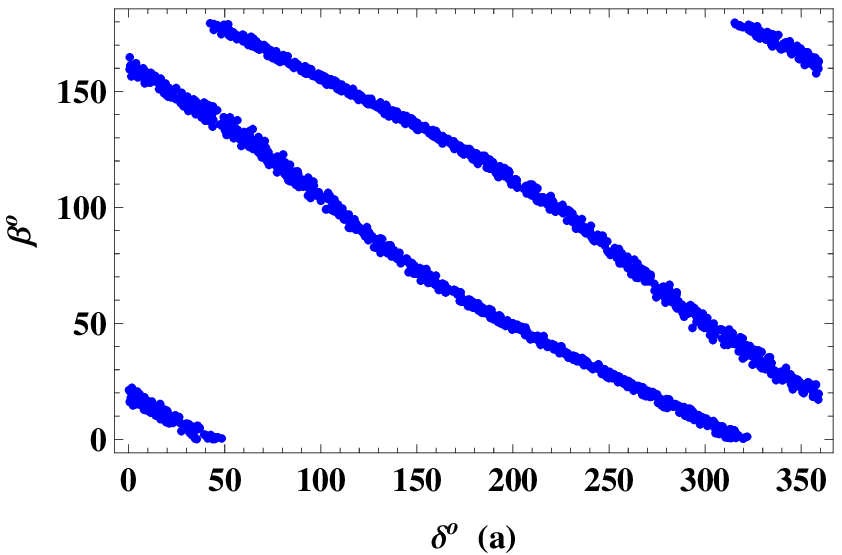,height=4.5cm,width=4.5cm}
\epsfig{file=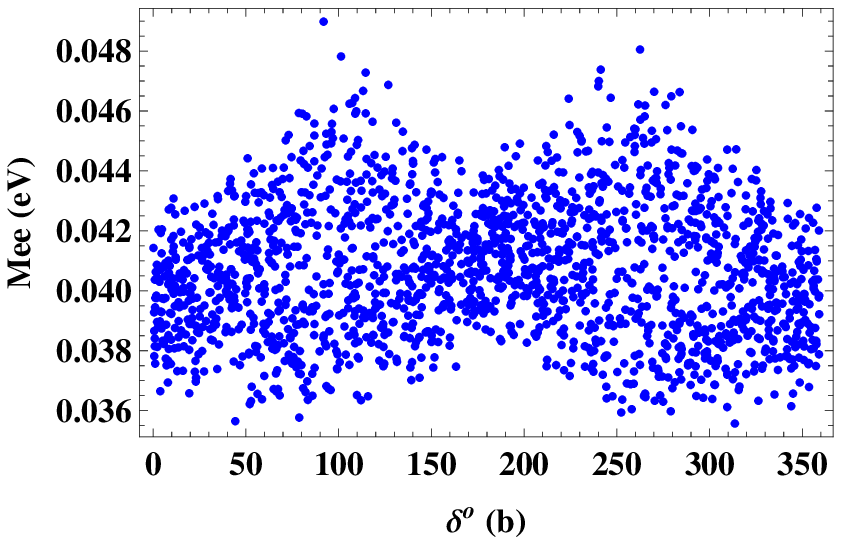,height=4.5cm,width=4.5cm}\\
\epsfig{file=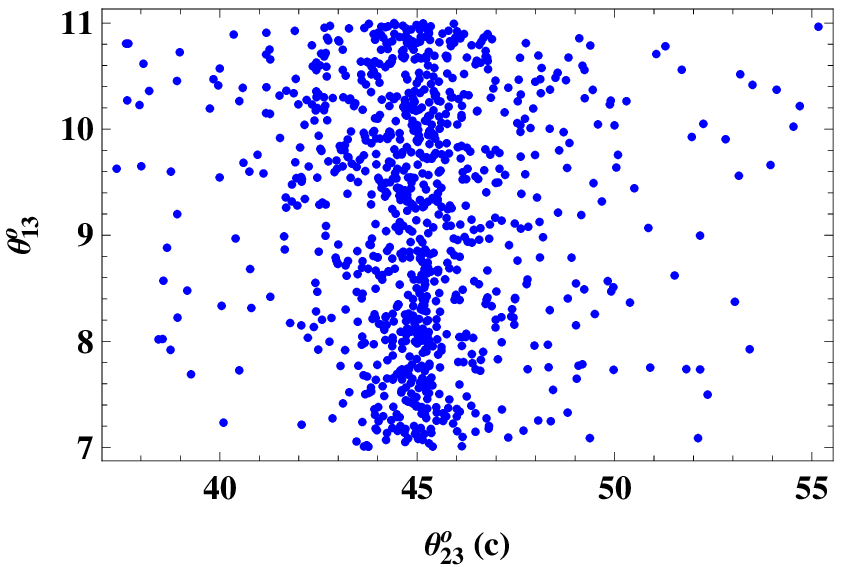,height=4.5cm,width=4.5cm}
\epsfig{file=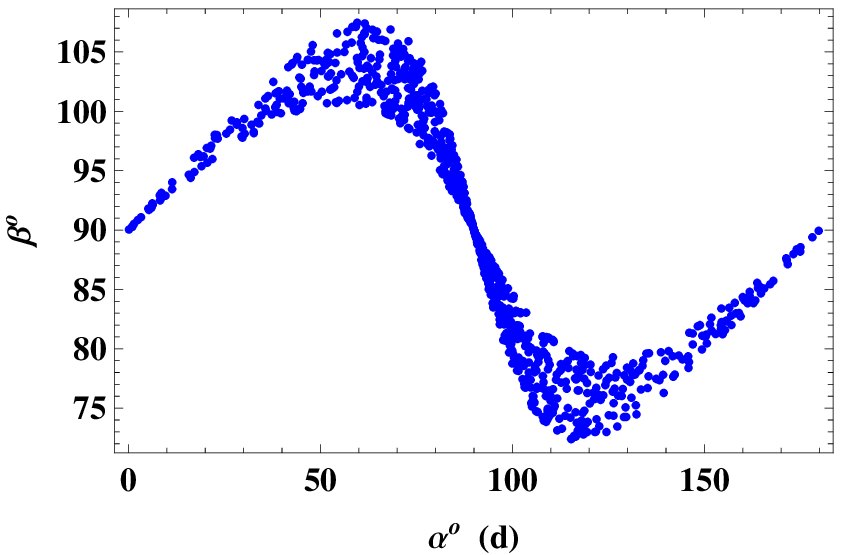,height=4.5cm,width=4.5cm}
\end{center}
\caption{The TEE correlation plots for classes $IXE(a)$(IS), $IXE(b)$(IS), $XF(c)$(NS) and $XID(d)$(IS).}
\end{figure} 
\begin{table}[!h]
\centering
\resizebox{5. in}{4.1 in}{\begin{tabular}{|c|c|c|c|c|}
 \hline 
Texture    & Spectrum   & $M_{ee}$ (eV)           & Lower bound on mass scale (eV)  & Majorana Phases         \\
 \hline
IA         & NS         & $0.004-0.12$            &  $0.001$         &  $\alpha = 0^{\circ}-70^{\circ},~110^{\circ}-180^{\circ}$\\
           & IS         & $0.02-0.16$             &  $0.001$         &  $\alpha = 50^{\circ}-70^{\circ},~110^{\circ}- 130^{\circ}$\\
 \hline 
IB (IVA)   & NS         & $0.006-0.14$            &  $0.007$         &  -                              \\
           & IS         & $0.02-0.18$             &  $0.007$         &  $\alpha = 40^{\circ}-140^{\circ}$  \\
 \hline
IC (ID)    & NS         & $0.02-0.12$             &  $0.023$         &  -              \\ 
           & IS         & $0.01-0.14$             &  $0.007$         &  $\alpha = 50^{\circ}-130^{\circ}$ \\
\hline  
IE (IIIC)  & NS         & $0.004-0.30$            &  $0.0030$        &  -                              \\
           & IS         & $0.01-0.30$             &  $0.0004$        &  -                              \\
\hline
IF (IIIF)  & NS         & $0.004-0.30$            &  $0.007$         &  -                              \\
           & IS         & $0.02-0.30$             &  $0.001$         &  -                              \\
 \hline
IIA (IIIB) & NS         & $0.007-0.12$            &  $0.006$         &  $\alpha = 30^{\circ}-80^{\circ},~100^{\circ}-150^{\circ}$ \\ 
           & IS         & $0.02-0.16$             &  $0.040$         &  $\alpha = 60^{\circ}-120^{\circ}$            \\
 \hline
IIB (IIIA) & NS         & $0.01-0.11$             &  $0.017$         &  $\alpha = 20^{\circ}-160^{\circ}$            \\ 
           & IS         & $0.02-0.16$             &  $0.020$         &  $\alpha = 40^{\circ}-130^{\circ}$           \\
  \hline
IIC        & NS         & $0.01-0.30$             &  $0.0100$        &  -                              \\
           & IS         & $0.01-0.22$             &  $0.0033$        &  $\alpha = 20^{\circ}-180^{\circ}$            \\
\hline 
IID        & NS         & $0.01-0.30$             & $0.01$           &  $\alpha = 0^{\circ}-80^{\circ},~100^{\circ}-180^{\circ}$ \\
           & IS         & $0.01-0.30$             & $0.01$           &  -                              \\
\hline 
IIE (VIIC) & NS         & $0.015-0.20$            & $0.02$           &  -                              \\   
           & IS         & $0.01-0.18$             & $0.02$           &  $\alpha = 20^{\circ}-160^{\circ}$  \\
\hline
IIF        & NS         & $0.01-0.30$             & $0.025$          &  -                              \\
           & IS         & $0.03-0.30$             & $0.004$          &  -                              \\
\hline 
IIID (VID) & NS         & $0.01-0.30$             & $0.025$          &  -                              \\
           & IS         & $0.01-0.30$             & $0.025$          &  -                              \\
\hline
IVB (VIIA) & NS         & $0.001-0.16$            & $0.0040$         &  $\alpha = 0^{\circ}-160^{\circ}$   \\
           & IS         & $0.02-0.16$             & $0.0005$         &  $\alpha =50^{\circ}-75^{\circ},~105^{\circ}-130^{\circ}$ \\
\hline  
IVC (VE)   & NS         & $0.005-0.25$            & $0.025$          &  -                              \\ 
           & IS         & $0.02-0.30$             & $0.030$          &  -                              \\
 \hline
IVE (VC)   & NS         & $0.01-0.18$             & $0.016$          &  -                              \\
           & IS         & $0.02-0.14$             & $0.016$          &   $\alpha=40^{\circ}-140^{\circ}$        \\
\hline
IVF (VIF)  & NS         & $0.003-0.30$            & $0.0033$         & $\alpha =0^{\circ}-20^{\circ},~160^{\circ}-180^{\circ}$ \\
           & IS         & $0.01-0.30$             & $0.00475$        &  -                              \\
 \hline
VA (VIB)   & NS         & $0.001-0.16$            & $0.005$          &  -                                \\
           & IS         & $0.02-0.14$             & $0.005$          &  $\alpha = 40^{\circ}-140^{\circ}$     \\
\hline
VB (VIA)   & NS         & $0.0012-0.09$           & $0.0165$         &  $\alpha = 40^{\circ}-140^{\circ}$      \\
\hline
VIE (VIIIC)& NS         & $0.006-0.18$            & $0.005$          &  -                              \\
           & IS         & $0.010-0.18$            & $0.0002$         &  $\alpha = 30^{\circ}-140^{\circ}$    \\
\hline
VIIB (XA)  & NS         & $0-0.12$                & $0.0001$         &  -                              \\
           & IS         & $0.02-0.12$             & $0.0002$         &  $\alpha = 40^{\circ}-140^{\circ}$\\
 \hline
VIID       & NS         & $0.01-0.30$             & $0.0100$         &  -                     \\
           & IS         & $0.01-0.30$             & $0.006$          &  -                              \\
\hline 
VIIE (IXC) & NS         & $0.005-0.16$            & $0.0300$         &  $\alpha = 50^{\circ}-130^{\circ}$             \\
           & IS         & $0.03-0.12$             & $0.0003$         &  $\alpha =20^{\circ}-80^{\circ},~100^{\circ}-160^{\circ}$ \\
\hline 
VIIF (XB)  & NS         & $0.002-0.14$            & $0.0140$         &  $\alpha = 40^{\circ}-160^{\circ}$  \\
           & IS         & $0.02-0.14$             & $0.0001$         &  $\alpha = 20^{\circ}-160^{\circ}$             \\
\hline 
VIIIA (IXB)& NS         & $0.0004-0.10$           & $0$              &  -                              \\
           & IS         & $0.02-0.14$             & $0.0001$         &  $\alpha = 40^{\circ}-140^{\circ}$      \\
\hline
VIIIB (IXA)& NS         & $0.001-0.16$            & $0.0001$         &  -                              \\
           & IS         & $0.02-0.16$             & $0.0010$         &  $\alpha =50^{\circ}-80^{\circ},~100^{\circ}-130^{\circ}$ \\
\hline 
VIIID      & NS         & $0.005-0.30$            & $0.025$          &  -                              \\
           & IS         & $0.028-0.30$            & $0.001$          &  -                              \\
\hline 
VIIIE (IXD)& NS         & $0.005-0.14$            & $0.0270$         &  $\alpha = 50^{\circ}-130^{\circ}$     \\
           & IS         & $0.03-0.12$             & $0.0002$         &  $\alpha = 20^{\circ}-80,~100^{\circ}-160^{\circ}$ \\
\hline
VIIIF (XC) & NS         & $0.005-0.30$            & $0.0200$         &  -                              \\ 
           & IS         & $0.02-0.30$             & $0.0003$         &  -                              \\
\hline
IXE (IXF)  & IS         & $0.03-0.05$             & $0.005$          &  $\alpha = 30^{\circ}-50^{\circ},~130^{\circ}-150^{\circ}$ \\
 \hline
XD (XIC)   & NS         & $0.002-0.18$            & $0.0010$         &  -            \\
           & IS         & $0.02-0.13$             & $0.0005$         &  $\alpha =0^{\circ}-70^{\circ},~110^{\circ}-180^{\circ}$   \\
\hline
XE (XIB)   & NS         & $0.002-0.30$            & $0.017$          &  -                              \\
           & IS         & $0.02-0.30$             & $0.001$          &  -                              \\
\hline  
XF         & NS         & $0.005-0.30$            & $0.025$          &  -                              \\
           & IS         & $0.025-0.30$            & $0.002$          &  $\alpha = 0^{\circ}-80^{\circ},~100^{\circ}-180^{\circ}$  \\
\hline  
XID        & IS         & $0.01-0.055$            & $0.0005$         &  $\beta = 70^{\circ}-110^{\circ}$             \\
\hline
XIE (XIF)  & NS         & $0.001-0.12$            & $0.0080$         &  -        \\
           & IS         & $0.01-0.18$             & $0.0002$         &  -        \\
 \hline    
\end{tabular}}
\caption{The numerical predictions for the phenomenologically viable textures in the case of two equalities between the elements of $M_\nu$.}
\end{table}

\begin{itemize}
\item Five textures, viz., \\
$IIIE$, $VIC$, $IVD$, $VD$ and $VF$ are excluded by the experimental data. 
\item Textures $IXE$, $IXF$ and $XID$ lead to an inverted spectrum only.
\item Textures $VB$ and $VIA$ satisfy normal spectrum only.
\item The allowed points for the following textures are very few:\\
$IA$, $IE$, $IIIC$, $IIA$, $IIIB$, $IID$ for inverted spectrum\\
$IIC$, $XE$, $XIB$, $VIIE$, $IXC$ for normal spectrum\\
$VIIF$, $XC$ for both inverted and normal spectrum.\\ 
We have generated $10^8$ points for these textures.
\item All the viable textures except $IA$(NS), $IXE$, $IXF$ and $XID$ allow quasi-degenerate spectrum.
\item The value of $M_{ee}$ is bounded from below for most of the viable textures.
\item It is found that the smallest neutrino mass cannot be zero for any of the allowed textures except for $VIIIA$(NS) and $IXB$(NS). 
\item For textures\\
$IE$, $IIIC$, $IF$, $IIIF$, $IIF$, $IVF$ and $VIF$, a non-vanishing reactor mixing angle is an inherent property since for $\theta_{13}=0$ these textures predict $m_1 = m_2$ which contradicts the experimental observations.
\end{itemize}

The numerical results for all the presently viable classes are summarised in Table 3. We have presented some of the interesting results in Figs. 1-3. Fig. 1(a) and 1(b) show the 2-3 interchange symmetry between classes $IF$ and $IIIF$. From Fig. 1(c), we can see that for IS in class $IIIB$, $\theta_{23}$ remains below maximal. Fig. 1(d) shows the correlation plot between $\delta$ and $M_{ee}$ for class $VB$(NS). Figs. 2(a) and 2(b) show the plots for class $IVF$(NS) which is one of the most predictive classes in this analysis. The predictions for $\delta$ strongly depend on the value of $M_{ee}$. Class $VIIID$  predicts $\theta_{23}$ to be near maximal as shown in Figs. 2(c) and 2(d). For class $IXE$(IS), there is a strong correlation between the Dirac-type phase $\delta$ and the Majorana-type phase $\beta$. Moreover, $M_{ee}$ is restricted to a very small range in this case [Figs. 3(a) and 3(b)]. For class $XF$(NS), $\theta_{23}$ near its maximal value is more probable [Fig. 3(c)]. As shown in Fig. 3(d), there is a strong correlation between the two Majorana-type CP-violating phases for class $XID$(IS).

\subsection{Numerical Results for TEC}
The numerical results for two equalities between the cofactors of the neutrino mass matrix are presented here. The main outcomes are: 
\begin{figure}[!]
\begin{center}
\epsfig{file=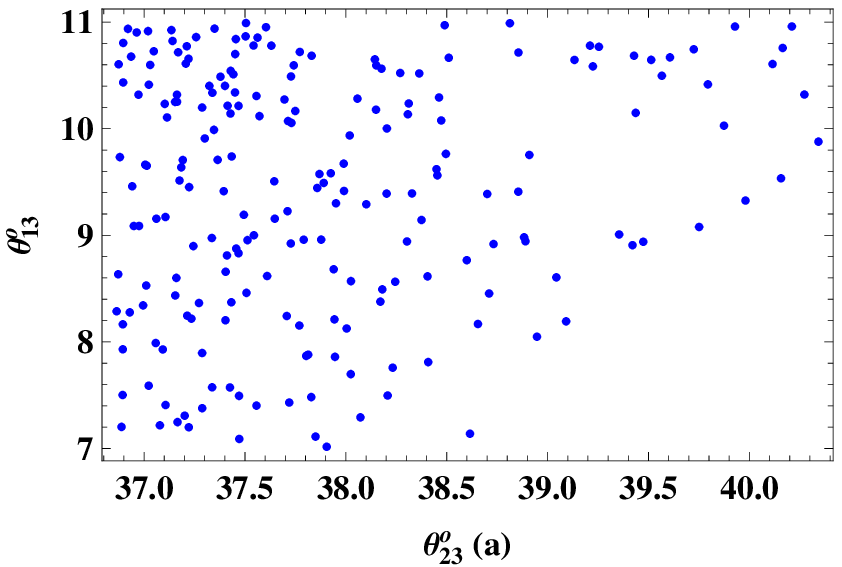,height=4.5cm,width=4.5cm}
\epsfig{file=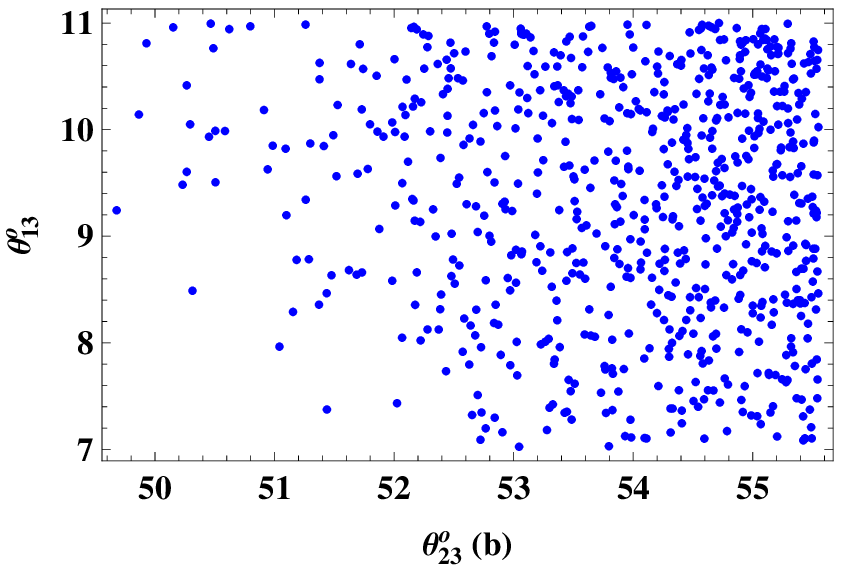,height=4.5cm,width=4.5cm}\\
\epsfig{file=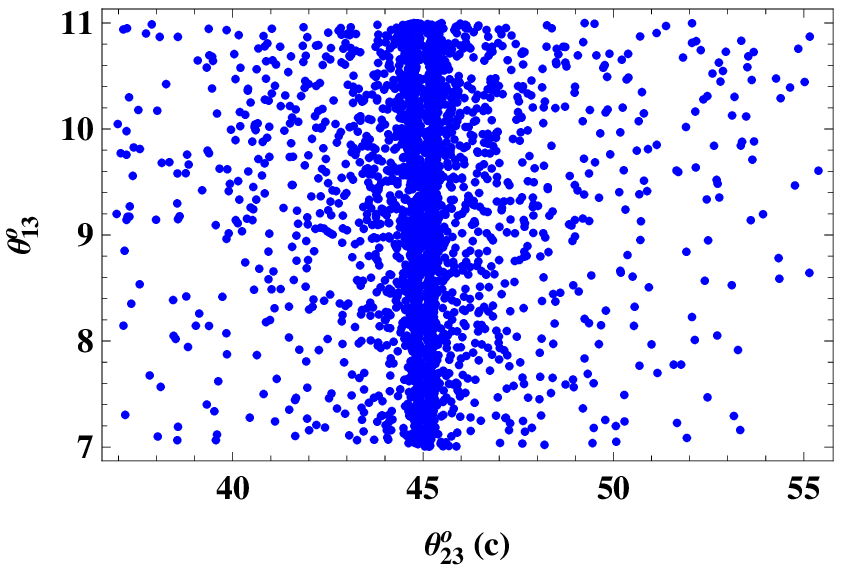,height=4.5cm,width=4.5cm}
\epsfig{file=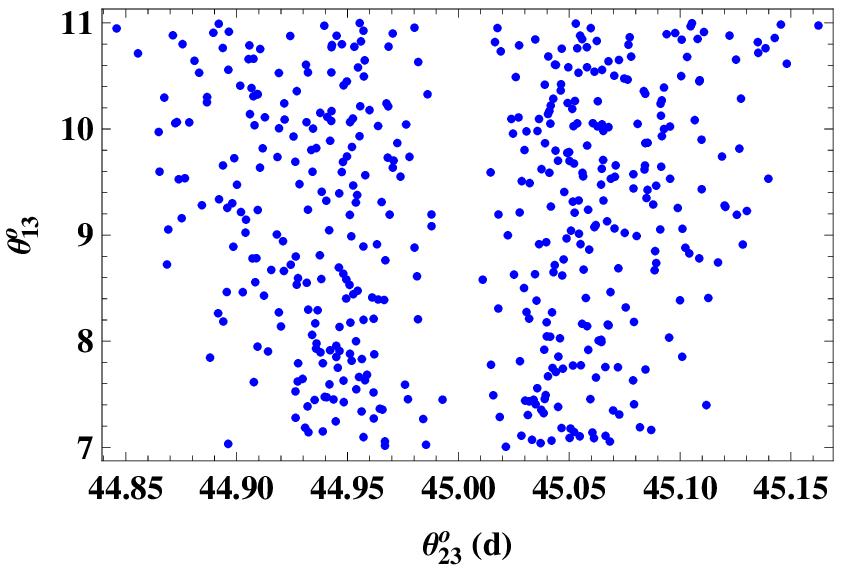,height=4.5cm,width=4.5cm}
\end{center}
\caption{The TEC correlation plots for classes $IIIB(a)$(NS), $IIA(b)$(NS), $VIIID(c)$(NS) and $VIIID(d)$(IS).}
\end{figure}

\begin{figure}[!]
\begin{center}
\epsfig{file=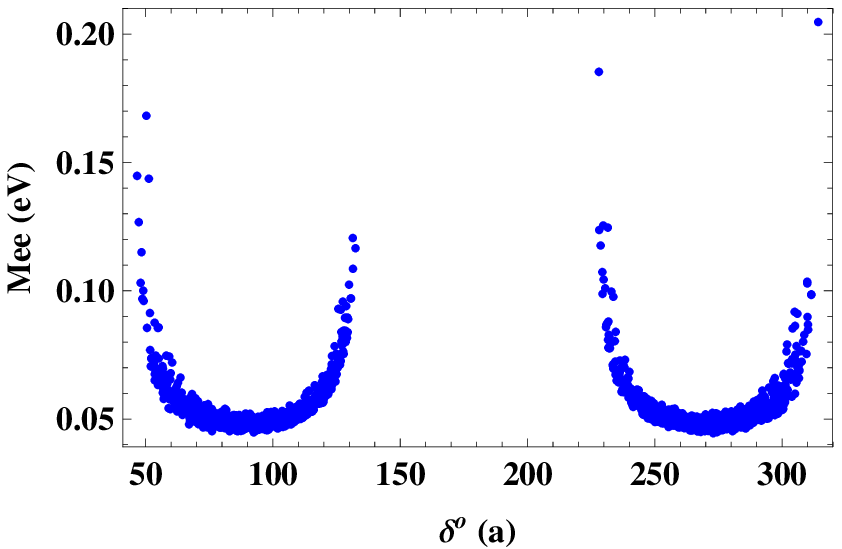,height=4.5cm,width=4.5cm}
\epsfig{file=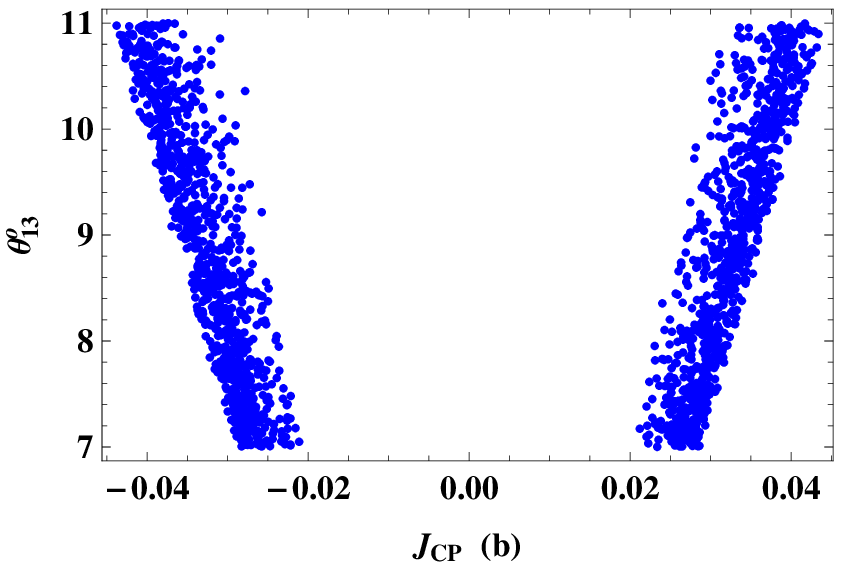,height=4.5cm,width=4.5cm}\\
\epsfig{file=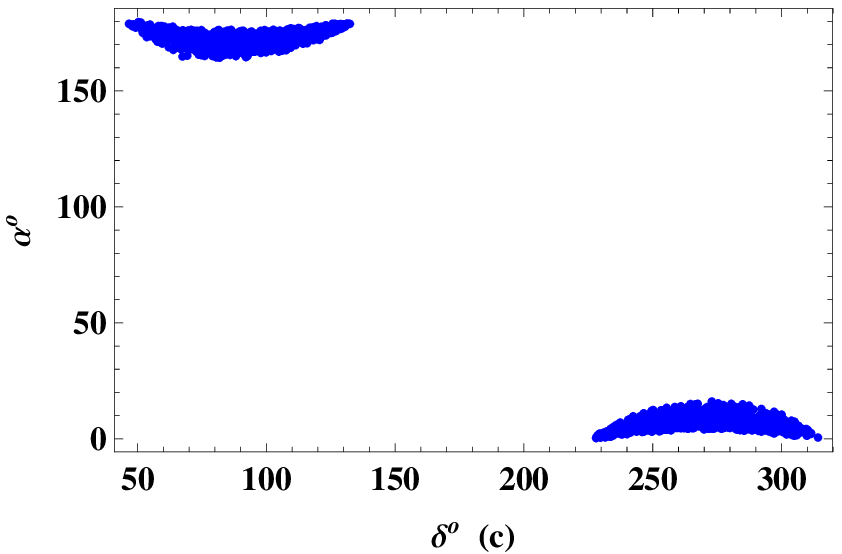,height=4.5cm,width=4.5cm}
\epsfig{file=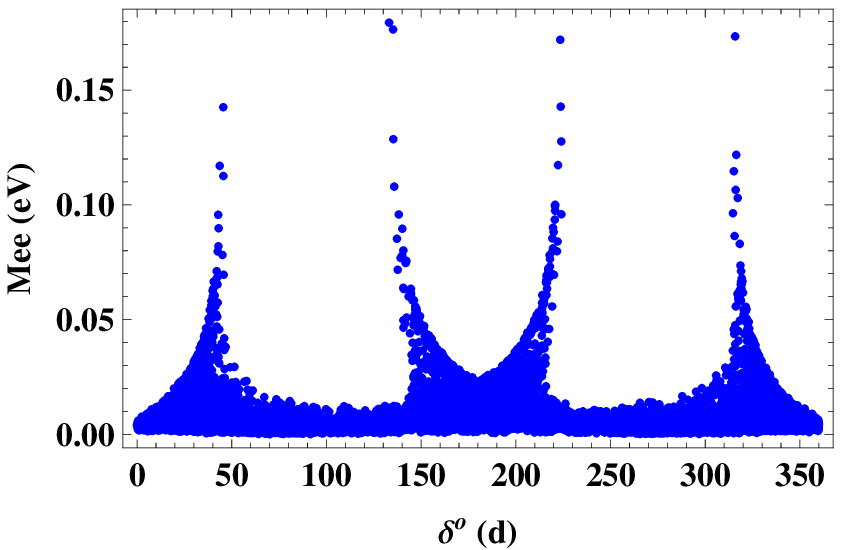,height=4.5cm,width=4.5cm}
\end{center}
\caption{The TEC correlation plots for classes $IVF(a, b, c)$(IS) and $IVF(d)$(NS).}
\end{figure}

\begin{figure}[!]
\begin{center}
\epsfig{file=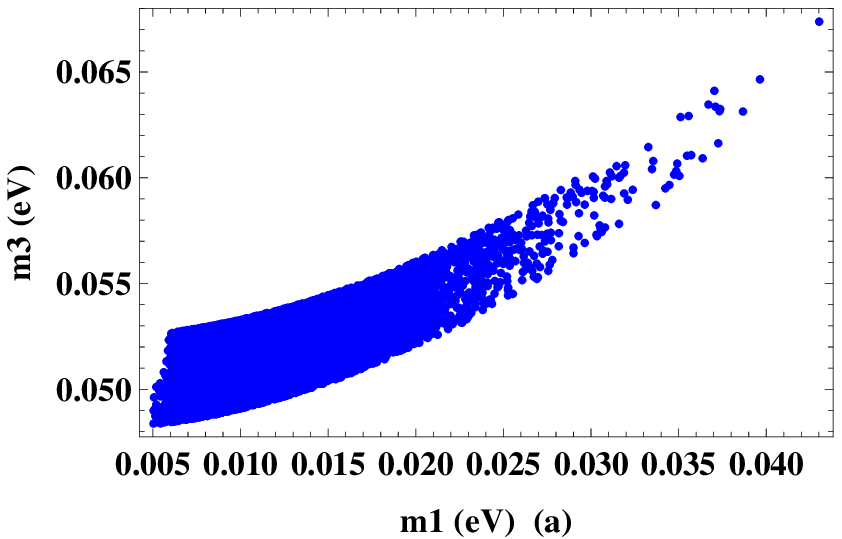,height=4.5cm,width=4.5cm}
\epsfig{file=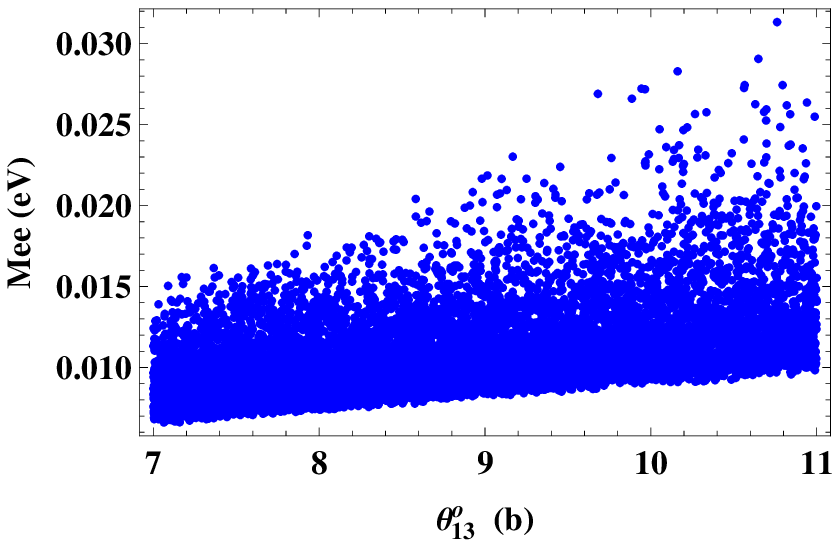,height=4.5cm,width=4.5cm}\\
\epsfig{file=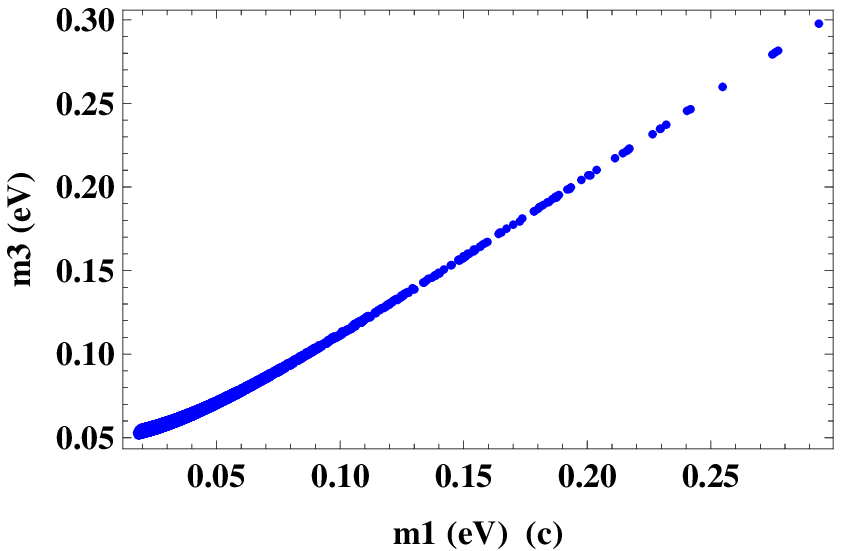,height=4.5cm,width=4.5cm}
\epsfig{file=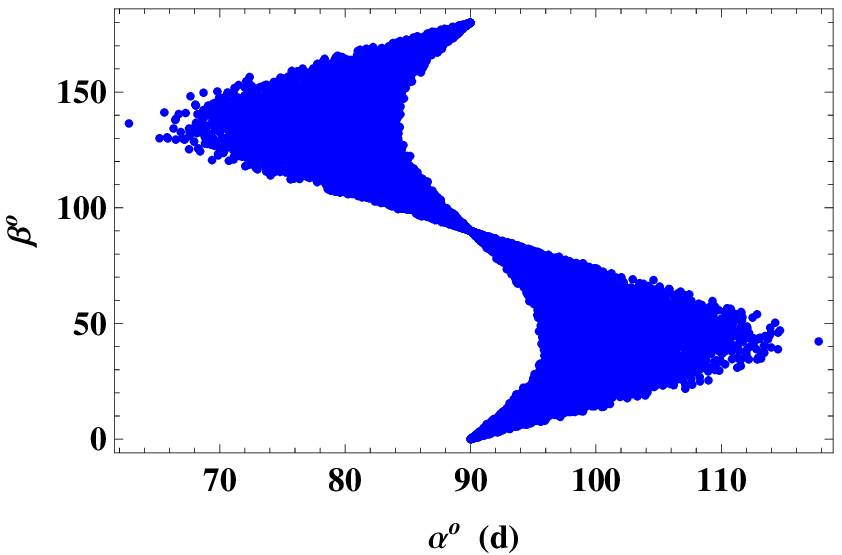,height=4.5cm,width=4.5cm}
\end{center}
\caption{The TEC correlation plots for classes $IXE(a)$(NS), $IXE(b)$(NS), $VIID(c)$(NS) and $XID(d)$(NS).}
\end{figure}
\begin{table}[!h]
\centering
\resizebox{5. in}{4.1 in}{\begin{tabular}{|c|c|c|c|c|}
 \hline 
Texture    & Spectrum   & $M_{ee}$ (eV)      & Lower bound on mass scale (eV)       &  Majorana Phases \\
 \hline
IA         & NS         & $0.0004- 0.07$     & $0.003$                      &  $\alpha= 50^{\circ}-130^{\circ}$  \\
           & IS         & $ 0.02-0.08 $      & $0.001$                      &  $\alpha= 0^{\circ}-70^{\circ},~110^{\circ}-180^{\circ}$ \\
\hline 
IB (IVA)   & NS         & $0.005-0.16$       & $0.015$                      &  $\alpha= 40^{\circ}-140^{\circ}$ \\ 
           & IS         & $0.015-0.14$       & $0.010$                      &  -                                  \\
 \hline
IC (ID)    & NS         & $0.001-0.14$       & $0.018$                      &  $\alpha = 50^{\circ}-130^{\circ}$ \\
           & IS         & $0.035-0.18$       & $0.020$                      &  -                                \\
\hline
IE (IIIC)  & NS         & $0.001-0.30$       & $0.010$                      &  -                                \\
           & IS         & $0.02-0.30$        & $0.006$                      &  -                                 \\
\hline
IF (IIIF)  & NS         & $0.003-0.30$       & $0.001$                      &  -                                 \\
           & IS         & $0.010-0.30$       & $0.010$                      &  -                                  \\
\hline
IIA (IIIB) & NS         & $0.02-0.16$        & $0.040$                      &  $\alpha = 60^{\circ}-120^{\circ}$     \\
           & IS         & $0.02-0.16$        & $0.006$                      &  $\alpha =40^{\circ}-80^{\circ},~100^{\circ}-140^{\circ}$ \\
 \hline
IIB (IIIA) & NS         & $0.01-0.12$        & $0.025$                      &  $\alpha = 50^{\circ}-140^{\circ}$     \\
           & IS         & $0.02-0.14$        & $0.018$                      &  $\alpha = 20^{\circ}-160^{\circ}$       \\
  \hline
IIC        & NS         & $0.003-0.30$       & $0.010$                      &  $\alpha = 20^{\circ}-160^{\circ}$     \\
           & IS         & $0.025-0.25$       & $0.010$                      &  -                                    \\
\hline 
IID        & NS         & $0.001-0.30$       & $0.015$                      &  -                                     \\
           & IS         & $0.030-0.30$       & $0.010$                      &  -                                     \\
\hline 
IIE (VIIC) & NS         & $0.004-0.12$       & $0.020$                      &  $\alpha = 20^{\circ}-160^{\circ}$       \\
           & IS         & $0.020-0.14$       & $0.020$                      &  -                                      \\
\hline
IIF        & NS         & $0.003-0.30$       & $0.001$                      &  -                                        \\
           & IS         & $0.010-0.30$       & $0.030$                      &  -                                      \\
\hline 
IIID (VID) & NS         & $0.005-0.25$       & $0.025$                      &  -                                       \\
           & IS         & $0.025-0.30$        & $0.025$                      &  -                                      \\
\hline
IVB (VIIA) & NS         & $0.001-0.16$       & $0.003$                      &  $\alpha = 50^{\circ}-80^{\circ},~100^{\circ}-130^{\circ}$ \\
           & IS         & $0.010-0.16$       & $0.008$                      &  $\alpha = 40^{\circ}-140^{\circ}$       \\
\hline  
IVC (VE)   & NS         & $0.01-0.30$        & $0.028$                      &  -                                        \\ 
           & IS         & $0.01-0.30$        & $0.028$                      &  -                                        \\
 \hline
IVE (VC)  & NS          & $0.007-0.16$       & $0.018$                      &  $\alpha = 40^{\circ}-140^{\circ}$       \\ 
          & IS          & $0.01-0.18$        & $0.020$                      &  -                                         \\
\hline
IVF (VIF) & NS          & $0-0.20$           & $0.0001$                     &  -                                         \\ 
          & IS          & $0.04-0.22$        & $0.0001$                     &  $\alpha = 0^{\circ}-20^{\circ},~160^{\circ}-180^{\circ}$\\
\hline
VA (VIB)  & NS          & $0.002-0.14$       & $0.004$                      &  $\alpha = 40^{\circ}-140^{\circ}$            \\
          & IS          & $0.02-0.16$        & $0.010$                      &  $\alpha = 30^{\circ}-150^{\circ}$           \\
\hline
VB (VIA)  & IS          & $0.01-0.08$        & $0.020$                      &  $\alpha = 40^{\circ}-140^{\circ}$            \\
\hline
VIE (VIIIC)& NS         & $0.0001-0.16$      & $0.010$                      &  $\alpha = 40^{\circ}-140^{\circ}$            \\
           & IS         & $0.01-0.14$        & $0.010$                      &  -                                             \\
\hline
VIIB (XA)  & NS         & $0-0.14$           & $0.002$                      &  $\alpha = 40^{\circ}-140^{\circ}$            \\
           & IS         & $0.01-0.16$        & $0.001$                      &  -                                              \\
 \hline
VIID       & NS         & $0.002-0.30$       & $0.015$                      &  -                                              \\
           & IS         & $0.025-0.30$       & $0.010$                      &  -                                              \\
\hline 
VIIE (IXC) & NS         & $0.001-0.16$       & $0.001$                      &  $\alpha= 0^{\circ}-80^{\circ},~100^{\circ}-180^{\circ}$\\                                                  
           & IS         & $0.01-0.18$        & $0.030$                      &  $\alpha = 50^{\circ}-130^{\circ}$               \\                                                                          
\hline 
VIIF (XB)  & NS         & $0.001-0.14$       & $0.003$                      &  -                                              \\ 
           & IS         & $0.010-0.18$       & $0.014$                      &  $\alpha = 20^{\circ}-160^{\circ}$               \\                            
\hline 
VIIIA (IXB)& NS         & $0-0.08$           & $0.001$                      &  $\alpha= 40^{\circ}-140^{\circ}$              \\
           & IS         & $0.01-0.16$        & $0.001$                      &  -                                               \\
\hline
VIIIB (IXA)& NS         & $0-0.12$           & $0.001$                      &  $\alpha = 50^{\circ}-130^{\circ}$               \\ 
           & IS         & $0.01-0.18$        & $0.002$                      &  -                                               \\
\hline 
VIIID      & NS         & $0.002-0.30$       & $0.002$                      &  -                                                \\
VIIID      & IS         & $0.015-0.30$       & $0.027$                      &  -                                                  \\
\hline 
VIIIE (IXD)& NS         & $0.001-0.14$       & $0.001$                      &  -                                              \\ 
           & IS         & $0.01-0.20$        & $0.025$                      &   $\alpha = 50^{\circ}-130^{\circ}$               \\
\hline
VIIIF (XC) & NS         & $0.002-0.30$       & $0.0014$                     &  -                                                 \\  
           & IS         & $0.01-0.30$        & $0.0200$                     &  -                                                 \\
\hline
IXE (IXF)  & NS         & $0.005-0.03$       & $0.004$                      &  $\alpha=20^{\circ}-50^{\circ},~130^{\circ}-160^{\circ}$ \\
 \hline
XD (XIC)   & NS         & $0.001-0.10$       & $0.0010$                     &  $\alpha \neq~  90^{\circ}$                       \\
           & IS         & $0.01-0.18$        & $0.0135$                     &  -                                                  \\
\hline
XE (XIB)   & NS         & $0.001-0.30$       & $0.001$                      &  -                                              \\ 
           & IS         & $0.01-0.30$        & $0.020$                      &  -                                               \\
\hline  
XF         & NS         & $0.002-0.30$       & $0.001$                      &  $\alpha=0^{\circ}-80^{\circ},~100^{\circ}-180^{\circ}$\\
           & IS         & $0.01-0.30$        & $0.025$                      &  -                                                \\
\hline  
XID        & NS         & $0$                & $0.0005$                     &  $\alpha = 60^{\circ}-120^{\circ}$                \\
\hline
XIE (XIF)  & NS         & $0.001-0.16$       & $0.001$                      &  -                                                  \\
           & IS         & $0.01-0.16$        & $0.010$                      &  $\alpha = 20^{\circ}-160^{\circ}$                                                   \\
 \hline    
\end{tabular}}
\caption{The numerical predictions for the phenomenologically viable textures in the case of two equalities between the cofactors of $M_\nu$.}
\end{table}

\begin{itemize}
\item Five textures, viz.,\\
$IIIE$, $VIC$, $IVD$, $VD$ and $VF$ are excluded by the experimental data. 
\item Textures $IXE$, $IXF$ and $XID$ lead to a normal spectrum only.
\item Textures $VB$ and $VIA$ hold for an inverted spectrum only.
\item The allowed points for the following textures are very few:\\
$IC$, $ID$, $IIC$, $IIID$, $VID$, $VIID$, $VIIE$, $IXC$, $VIIF$, $XB$, $VIIIB$, $IXA$, $VIIIE$, $IXD$, $VIIIF$, $XC$, $XD$, $XIC$, $XE$, $XIB$, $XIE$, $XIF$ for inverted spectrum \\
$IIA$, $IIIB$ for normal spectrum\\
$IVC$, $VE$ for both inverted and normal spectrum. \\
We have generated $10^8$ points for these textures.
\item All the viable textures except $IXE$, $IXF$ and $XID$ allow quasi-degenerate spectrum.
\item For most of the viable textures, we obtain a lower bound on $M_{ee}$, but, for texture $XID$ this parameter can only have a vanishing value.
\item It is found that the smallest neutrino mass cannot be zero for any of the allowed textures.
\item For textures:\\
$IE$, $IIIC$, $IF$, $IIIF$, $IIF$, $IVF$ and  $VIF$, a non-vanishing reactor mixing angle is an inherent property.

\end{itemize}
The numerical results for all the classes which satisfy the present experimental data are summarised in Table 4. Some of the interesting results are plotted in Figs. 4-6. Figs. 4(a) and 4(b) show correlation plots for Classes $IIIB$ and $IIA$(NS) which are related by 2-3 interchange symmetry as the value of $\theta_{23}$ is below maximal for class $IIIB$ whereas for class $IIA$ it is above maximal. Figs. 4(c) and 4(d) correspond to class $VIIID$ for NS and IS respectively and the atmospheric mixing angle $\theta_{23}$ is restricted to the proximity of its maximal value for IS. In Fig. 5, we have depicted the correlation plots for class $IVF$ which is one of the most predictive classes of TEC, Figs. 5(a, b, c) correspond to IS while Fig. 5(d) corresponds to NS. The Dirac-type phase $\delta$ is correlated with $M_{ee}$ [Fig. 5(a)]. It is clear from Fig. 5(b) that this class is necessarily CP-violating because the Jarlskog CP-violation rephasing invariant cannot vanish in this case. Figs. 6(a) and 6(b) correspond to class $IXE$(NS) for which the unknown parameter $M_{ee}$ is restricted to a very small range. In Fig. 6(c), we have plotted $m_1$ and $m_3$ for class $VIID$(NS) for which the quasidegenerate limit is allowed as is the case for most of the classes included in this analysis. The two Majorana-type CP-violating phases $\alpha$ and $\beta$ for class $XID$ are plotted in Fig. 6(d). For this class, only normal spectrum is allowed and the unknown parameter $M_{ee}$ can only have a vanishing value because the $M_R$ corresponding to this texture has a vanishing cofactor corresponding to the $(e,e)$ entry which manifests as a vanishing $M_{ee}$ in the neutrino mass matrix $M_\nu$.\\ Comparing the numerical results of TEE and TEC, we find that the phenomenological predictions for a texture are in general similar for both cases (\textit{i.e.} TEE and TEC) except for mass hierarchy which gets reversed. For example, the disallowed classes are the same in both cases. The textures which only satisfy NS in the case of TEE get replaced by IS for TEC and vice versa. The textures for which a non-vanishing reactor mixing angle is an inherent property are the same in both cases. Thus the distinguishing feature between TEE and TEC for a texture, in general, is the neutrino mass hierarchy. The reason for similar phenomenological predictions (and for reversal of mass hierarchy) of corresponding textures of TEE and TEC can be understood in the following way.\\ 
The diagonalization equation of $M_\nu$ [Eq. (2)] is
\begin{equation}
M_{\nu}= V' M_{\nu}^{diag}V'^{T}.
\end{equation}
Taking the inverse of $M_\nu$ gives
\begin{equation}
M_{\nu}^{-1}= V'^* (M_{\nu}^{diag})^{-1}V'^{\dagger}
\end{equation}
For two equalities between the elements of $M_\nu$, $M_\nu^{-1}$ has two equalities between the cofactors and vice versa. Thus, TEE and TEC textures are just the inverse of each other. It can be seen from the above equations that the mixing matrices are complex conjugates of each other. Thus, they span the same parameter space for mixing angles and it is expected that the corresponding textures of TEC and TEE have similar phenomenological predictions for the mixing angles. However, it is not necessary that the allowed values of mixing angles are exactly the same in both cases because the mass eigenvalues are inversely related and different regions of mixing angles may be allowed from the whole parameter space when we use the input of mass squared differences.

\section{Symmetry Realization}
Here we present a simple $A_4 \times Z_2$ model for one of the cases, viz., $IIC$ of TEC studied in this analysis. All the leptonic fields are assigned to the triplet representation of $A_4$. The transformations of various fields are given in Table 5. These transformation properties lead to the following $A_4 \times Z_2$ invariant Lagrangian for the leptons:
\begin{eqnarray}
\mathcal{L} = Y_1 (\overline{D}_{e_L} e_R + \overline{D}_{\mu_L} \mu_R + \overline{D}_{\tau_L} \tau_R)_{\underline{\bf{1}}} \phi_1 + Y_2 (\overline{D}_{e_L} e_R + \omega \overline{D}_{\mu_L} \mu_R + \omega^2 \overline{D}_{\tau_L} \tau_R)_{\underline{\bf{1}^\prime}} \phi_3 + \nonumber \\ Y_3 (\overline{D}_{e_L} e_R + \omega^2 \overline{D}_{\mu_L} \mu_R + \omega \overline{D}_{\tau_L} \tau_R)_{\underline{\bf{1}^{\prime \prime}}} \phi_2 + Y_4 (\overline{D}_{e_L} \nu_{e_R} + \overline{D}_{\mu_L} \nu_{\mu_R} + \overline{D}_{\tau_L} \nu_{\tau_R})_{\underline{\bf{1}}} \tilde{\phi_4} + \nonumber \\ \frac{M_M}{2}( \nu_{e_R}^T C^{-1} \nu_{e_R} + \nu_{\mu_R}^T C^{-1} \nu_{\mu_R} + \nu_{\tau_R}^T C^{-1} \nu_{\tau_R})_{\underline{\bf{1}}} + \frac{Y_{M_1}}{2} [(\nu_{\mu_R}^T C^{-1} \nu_{\tau_R} + \nu_{\tau_R}^T C^{-1} \nu_{\mu_R}) \chi_1 \nonumber \\ + (\nu_{\tau_R}^T C^{-1} \nu_{e_R} + \nu_{e_R}^T C^{-1} \nu_{\tau_R}) \chi_2 + (\nu_{e_R}^T C^{-1} \nu_{\mu_R} + \nu_{\mu_R}^T C^{-1} \nu_{e_R}) \chi_3]_{\underline{\bf{1}}} + \ \ \textrm{H. c.}
\end{eqnarray}
where $\tilde{\phi_4} = i \tau_2 \phi_4^*$.
 The $Z_2$ symmetry is used to prevent the coupling of Higgs doublet $\phi_4$ with the charged leptons so that it only contributes to the Dirac neutrino mass matrix and vice-versa. When the various Higgs fields acquire non-zero vacuum expectation values (VEVs), the $A_4 \times Z_2$ invariant Yukawa Lagrangian leads to a diagonal charged lepton mass matrix
 \begin{equation}
 M_l = \left(
\begin{array}{ccc}
m_e & 0 & 0 \\ 0 & m_\mu & 0 \\ 0& 0 & m_\tau
\end{array}
\right)
 \end{equation}
where $m_e = Y_1 \langle \phi_1 \rangle_o + Y_2 \langle \phi_3 \rangle_o +Y_3 \langle \phi_2 \rangle_o$, $m_\mu = Y_1 \langle \phi_1 \rangle_o + \omega Y_2 \langle \phi_3 \rangle_o + \omega^2 Y_3 \langle \phi_2 \rangle_o$ and $m_\tau = Y_1 \langle \phi_1 \rangle_o + \omega^2 Y_2 \langle \phi_3 \rangle_o + \omega Y_3 \langle \phi_2 \rangle_o$. The Dirac neutrino mass matrix is proportional to the $3 \times 3$ identity matrix
 \begin{equation}
 M_D = Y_4 \langle \Phi_4 \rangle_o \ \textbf{I}.
 \end{equation}
 The right handed Majorana mass matrix $M_R$ has the form
 \begin{equation}
 M_R = \left(
\begin{array}{ccc}
a & b & c \\ b & a & d \\ c& d & a
\end{array}
\right).
\end{equation}
The diagonal entries in $M_R$ come from the bare Majorana mass term and the off-diagonal entries arise via the Yukawa couplings with $\chi_i$. Since the off-diagonal elements of $M_R$ are supposed to be all different, the scalar potential must be rich enough for the VEVs of $\chi_i$ to be all different.
After the type-I seesaw, the effective neutrino mass matrix $M_\nu$ has TEC corresponding to the 11, 22 and 33 entries.  
\begin{table}[h]
\begin{center}
\begin{tabular}{|c|c|c|c|c|c|c|c|c|}
\hline Fields & $D_{l_{L}}$ & $l_R$ & $\nu_{l_{R}}$ & $\phi_1$ & $\phi_2$ & $\phi_3$ & $\phi_4$ & $\chi_i$ \\ 
\hline $SU(2)_L$ & 2 & 1 & 1 & 2 & 2 & 2 & 2 & 1 \\ 
\hline $A_4$ & $\bf{3}$ & $\bf{3}$ & $\bf{3}$ & $\bf{1}$ & $\bf{1^{\prime}}$& $\bf{1^{\prime \prime}}$ & $\bf{1}$ & $\bf{3}$ \\ 
\hline $Z_2$ & + & + & $-$ & + & + & + & $-$ & + \\ 
\hline 
\end{tabular}
\end{center}
\caption{Transformation properties of various fields for case $IIC$ of TEC.}
\end{table} 

For class $IIC$ of TEE, the transformations of various fields are given in Table 6. We have pointed out earlier that such textures in $M_\nu$ may arise through type-II seesaw for which we need to add four Higgs $SU(2)_L$ triplets $\triangle_i$ and no right-handed neutrino fields are needed. Three of the $SU(2)_L$ triplets transform in combination as $A_4$ triplet and fourth one is a singlet of $A_4$. A diagonal charged lepton mass matrix arises exactly in the same way as in the case $IIC$ of TEC, since the transformation properties of charged lepton fields remain the same as in the earlier case.\\ The $A_4$ invariant Lagrangian for neutrinos is 
\begin{small}
\begin{align}
\mathcal{L} =  & \frac{M_{L_1}}{2} [(D_{\mu_L}^T C^{-1} i \tau_2 \triangle_1 D_{\tau_L} + D_{\tau_L}^T C^{-1} i \tau_2 \triangle_1 D_{\mu_L}) + (D_{\tau_L}^T C^{-1} i \tau_2 \triangle_2 D_{e_L} + D_{e_L}^T C^{-1} i \tau_2 \triangle_2 D_{\tau_L}) + \nonumber \\ & (D_{e_L}^T C^{-1} i \tau_2 \triangle_3 D_{\mu_L} +  D_{\mu_L}^T C^{-1} i \tau_2 \triangle_3 D_{e_L})]_{\underline{\bf{1}}} + \frac{M_{L_2}}{2}(D_{e_L}^T C^{-1} i \tau_2 \triangle_4 D_{e_L} +  D_{\mu_L}^T C^{-1} i \tau_2 \triangle_4 D_{\mu_L} + \nonumber \\ & D_{\tau_L}^T C^{-1} i \tau_2 \triangle_4 D_{\tau_L})_{\underline{\bf{1}}} + \ \ \textrm{H. c.}
\end{align}
\end{small}
where the $SU(2)_L$ triplets are written in $2\times2$ matrix notation:
\begin{eqnarray}
\triangle = \left(
\begin{array}{cc}
H^+& \sqrt{2}H^{++}  \\
\sqrt{2}H^{0} &-H^+
\end{array}
\right).
\end{eqnarray}
When the neutral components of the $SU(2)_L$ triplet Higgs fields acquire small but non-zero and distinct VEVs, we get the neutrino mass matrix having the form $IIC$ of TEE
\begin{equation}
M_\nu = \left(
\begin{array}{ccc}
a & b & c \\ b & a & d \\ c& d & a
\end{array}
\right).
\end{equation}

\begin{table}[h]
\begin{center}
\begin{tabular}{|c|c|c|c|c|c|c|c|c|}
\hline Fields & $D_{l_{L}}$ & $l_R$ & $\nu_{l_{R}}$ & $\phi_1$ & $\phi_2$ & $\phi_3$ & $\triangle_i$ & $\triangle_4$ \\ 
\hline $SU(2)_L$ & 2 & 1 & 1 & 2 & 2 & 2 &3 & 3 \\ 
\hline $A_4$ & $\bf{3}$ & $\bf{3}$ & $\bf{3}$ & $\bf{1}$ & $\bf{1^{\prime}}$& $\bf{1^{\prime \prime}}$ & $\bf{3}$ & $\bf{1}$ \\ 
\hline 
\end{tabular}
\end{center}
\caption{Transformation properties of various fields for case $IIC$ of TEE.}
\end{table} 

\section{Summary}
We studied in detail the implications of the presence of two equalities between the elements or the cofactors of the neutrino mass matrix. Two equalities between the elements of the neutrino mass matrix can be obtained through the type-II seesaw mechanism whereas two equalities between the cofactors are obtained through the type-I seesaw mechanism when the right handed neutrino mass matrix has two equalities between elements and the Dirac neutrino mass matrix is proportional to the unit matrix. A total of sixty five texture structures are possible for each case. Predictions for $M_{ee}$ are given for the allowed texture structures. This parameter is expected to be measured in the forthcoming NDB decay experiments. To illustrate how such texture structures can be realised, we presented a simple $A_4$ model for the texture structure $IIC$. The viability of these textures suggests that there are still rich unexplored structures of the neutrino mass matrix from both the phenomenological and theoretical points of view. Future data from various experiments along with the input of flavor symmetry will help in deciding the form of the neutrino mass matrix.\\ 

\textbf{\textit{\Large{Acknowledgements}}}\\
The research work of S. D. and L. S. is supported by the University Grants
Commission, Government of India \textit{vide} Grant No. 34-32/2008
(SR). R. R. G. acknowledges the financial support provided by the Council for Scientific and Industrial Research (CSIR), Government of India.

\renewcommand{\theequation}{A-\arabic{equation}}
\setcounter{equation}{0}
\section*{A \ \ The Group $A_4$}
$A_4$ is the group of even permutations of four objects having twelve elements. Geometrically, it can be viewed as the group of rotational symmetries of the tetrahedron. $A_4$ has four inequivalent irreducible representations (IRs) which are three singlets $\textbf{1}$, $\textbf{1}^\prime$ and $\textbf{1}^{\prime\prime}$, and one triplet \textbf{3}. $A_4$ is generated by two generators $S$ and $T$ such that 
\begin{equation}
S^2 = T^3 = (S T)^3 = 1.
\end{equation}
The one dimensional unitary IRs are
\begin{align}
\textbf{1} \ \ \ \ \ &  S = 1 \ \ \ \ \ T = 1 \ ,\nonumber \\
\textbf{1}^{\prime} \ \ \ \ \ &  S = 1 \ \ \ \ \ T = \omega \ , \\
\textbf{1}^{\prime\prime} \ \ \ \ \ & S = 1 \ \ \ \ \ T = \omega^2 \ . \nonumber
\end{align}
The three dimensional unitary IR is 
 \begin{equation}
 S = \left(
\begin{array}{ccc}
1 & 0 & 0 \\ 0 & -1 & 0 \\ 0& 0 & -1
\end{array}
\right), \ \ T = \left(
\begin{array}{ccc}
0 & 1 & 0 \\ 0 & 0 & 1 \\ 1& 0 & 0
\end{array}
\right).
\end{equation}
The multiplication rules of the IRs are as follows
\begin{equation}
\textbf{1}^\prime \otimes \textbf{1}^\prime = \textbf{1}^{\prime \prime}, \ \textbf{1}^{\prime \prime} \otimes \textbf{1}^{\prime \prime} = \textbf{1}^{\prime}, \ \textbf{1}^{\prime} \otimes \textbf{1}^{\prime \prime} = \textbf{1}.
\end{equation}
The product of two $\textbf{3}$'s gives 
\begin{equation}
\textbf{3} \otimes \textbf{3} = \textbf{1} \oplus \textbf{1}^\prime \oplus \textbf{1}^{\prime \prime} \oplus \textbf{3}_s \oplus \textbf{3}_a
\end{equation}
where s(a) denotes symmetric(anti-symmetric) product. Let $(x_1, x_2, x_3)$ and $(y_1, y_2, y_3)$ denote the basis vectors of two $\textbf{3}$'s. Then the IRs obtained from their products are
\begin{align}
(\textbf{3} \otimes \textbf{3})_{\textbf{1}} & = x_1 y_1 + x_2 y_2 + x_3 y_3 \\
(\textbf{3} \otimes \textbf{3})_{\textbf{1}^\prime} & = x_1 y_1 + \omega x_2 y_2 + \omega^2 x_3 y_3 \\
(\textbf{3} \otimes \textbf{3})_{\textbf{1}^{\prime \prime}} & = x_1 y_1 + \omega^2 x_2 y_2 + \omega x_3 y_3 \\
(\textbf{3} \otimes \textbf{3})_{\textbf{3}_s} & = (x_2 y_3 + x_3 y_2, x_3 y_1 + x_1 y_3, x_1 y_2 + x_2 y_1) \\
(\textbf{3} \otimes \textbf{3})_{\textbf{3}_a} & = (x_2 y_3 - x_3 y_2, x_3 y_1 - x_1 y_3, x_1 y_2 - x_2 y_1).
\end{align}

\end{document}